\documentclass[floats,aps]{revtex4}

\usepackage{amsmath,latexsym,epsf}
\usepackage{epsfig,amsfonts}
\usepackage{amssymb}

\def\be{\begin{equation}}
\def\ee{\end{equation}}

\def\bea{\begin{eqnarray}}
\def\eea{\end{eqnarray}}

\def\e{\epsilon}

\def\s{\sigma}

\def\p{\varphi}
\def\t0{\tau_0}
\def\Tr{{\rm Tr}\,}
\def\b{\beta}
\def\l{\lambda}

\begin{document}

%
\title{Quantum Quenches in Extended Systems}

\author{Pasquale Calabrese${}^{1}$ and John Cardy$^2$}
\address{$^1$
Dipartimento di Fisica dell'Universit\`a di Pisa and INFN, Pisa, Italy\\
Institute for Theoretical Physics, University of Amsterdam,
1018 XE Amsterdam, The Netherlands.
}
\address{$^{2}$  Oxford University, Rudolf Peierls Centre for
Theoretical Physics, 
Oxford,
United Kingdom\\
All Souls College, Oxford, United Kingdom}
\date{\today}

\begin{abstract}

We study in general the time-evolution of correlation functions in
a extended quantum system after the quench of a parameter in the
hamiltonian. We show that correlation functions in $d$ dimensions
can be extracted using methods of boundary critical phenomena in
$d+1$ dimensions. For $d=1$ this allows to use the powerful tools
of conformal field theory in the case of critical evolution.
Several results are obtained in generic dimension in the gaussian
(mean-field) approximation. These predictions are checked against
the real-time evolution of some solvable models that allows also
to understand which features are valid beyond the critical
evolution.

All our findings may be explained in terms of a picture generally
valid, whereby quasiparticles, entangled over regions of the order
of the correlation length in the initial state, then propagate
with a finite speed through the system. Furthermore we show that
the long-time results can be interpreted in terms of a generalized
Gibbs ensemble. We discuss some open questions and possible future
developments.
\end{abstract}

\maketitle


\section{Introduction}

Suppose that an extended quantum system in $d$ dimensions (for
example a quantum spin system), is prepared at time $t=0$ in a
pure state $|\psi_0\rangle$ which is the ground state of some
hamiltonian $H_0$ (or, more generally, in a thermal state at a
temperature less than the mass gap to the first excited state).
For times $t>0$ the system evolves \em unitarily \em according to
the dynamics given by a \em different \em hamiltonian $H$, which
may be related to $H_0$ by varying a parameter such as an external
field. This variation, or quench, is supposed to be carried out
over a time scale much less than the inverse mass gap. How does
the state $|\psi(t)\rangle=e^{-iHt} |\psi_0\rangle$ evolve? For a
finite number of degrees of freedom, the system generically shows
a periodic (or quasiperiodic) behavior with a period that
typically increases when the number of degrees of freedom grows.
This is the well-known phenomenon of \em quantum recurrence\em.
However, in the thermodynamic limit this is no longer necessarily
the case, and the natural question arises as to whether the system
(or, rather, a macroscopically large subsystem) reaches a
stationary state for very large times. To attack this question we
consider the simpler question of how the correlation functions,
expectation values of products of local observables, evolve and
whether they reach constant values for large times.

This problem has its own theoretical interest, being a first step
towards the understanding of equilibration in quantum systems. On
the same fundamental level we can also ask whether the approach
can provide a new tool for the characterization of the collective
excitations in strongly correlated systems. However, until
recently it has been considered a largely academic question,
because the time scales over which most condensed matter systems
can evolve coherently without coupling to the local environment
are far too short, and the effects of dissipation and noise are
inescapable. Recent developments of experimental tools for
studying the behavior of ultra-cold atoms have revised completely
this negative attitude. In fact, thanks to the phenomenon of
Feshbach resonance it is possible to tune the coupling of an
interacting system to essentially any value and in extremely short
times. In addition, the coupling to dissipative degrees of freedom
can be much weaker than in ordinary condensed matter systems. This
has allowed the observation, among other things, of the collapse
and revival of a Bose-Einstein condensate \cite{uc} and of the
quenching of a spinor condensate \cite{spinor}. Furthermore, using
highly anisotropic optical lattices, it has been possible to build
and study essentially one-dimensional systems \cite{1dol}, and, of
direct interest for this paper, the coherent non-equilibrium
dynamics of these integrable models has been measured
\cite{kww-06}. (Other interesting experiments are considered in
Refs. \cite{uc2}.) These striking experimental results have
largely motivated the development of new numerical methods to
study non-equilibrium dynamics, the most successful one being the
time-dependent density matrix renormalization group (DMRG)
\cite{tdmrg}.

On the purely theoretical side these kind of questions were first
considered (as far as we are aware) in the seventies in the
context of the quantum Ising-XY model in Refs.~\cite{bm1,bm2} (see
also \cite{ir-00,ao-03,sps-04,cl-05,dssc-06,cc-07} for recent
developments). However, in the very last few years, after the
previously mentioned experimental progress, the study of quantum
quenches has been pursued in a systematic manner and a large
number of results is nowadays available. To quote only a few
examples, the models considered include several realizations of
one-dimensional (1D) Bose gases
\cite{gg,bilbao,rmrnm-06,rmo-06,topk-05,bbis-04,mg-05,ksdz-05,kla,cdeo-06,aa-02},
Luttinger liquids \cite{c-06,p-06}, coupled 1D condensates
\cite{gdlp-07}, strongly correlated 1D fermions \cite{mwnm-06},
and mean field fermion condensates \cite{bcs,df}. A closely
related topic, that will not be considered here, concerns the
formation of defects when crossing a critical point with changing
the external parameter at a fixed rate, i.e. the so-called quantum
Kibble-Zurek mechanism \cite{qzm}.

Most of these papers concern the exact or approximate solutions of
very specific models. These are often strongly relevant, since in
many cases they can be directly compared with existing
experimental results or give very accurate predictions for future
investigations. However, from the study of specific models it is
difficult to draw general conclusions on the physics of quantum
quenches. There have been at least two notable exceptions. In Ref.
\cite{gg} it was conjectured that the asymptotic state at very
large times can be described by a generalized Gibbs ensemble (for
more details we refer the reader to section \ref{phint} where this
conjecture will be explicitly discussed in relation to some of our
findings). In Ref.~\cite{cc-06} we studied the time evolution
after a quench in general, exploiting the path integral approach
and the well-known mapping of the quantum problem to a classical
one in $d+1$ dimensions. The translational invariance in the
(imaginary) time direction is explicitly broken and thus the
initial state plays the role of a boundary condition. This limits
the range of applicability of field-theoretical methods, but when
the hamiltonian $H$ is at or close to a quantum critical point we
could use the renormalization group (RG) theory of boundary
critical behavior (see, e.g., \cite{dd}). From this point of view,
particularly powerful analytic results are available for $d=1$
because then the $1+1$-dimensional problem is described
asymptotically by a boundary conformal field theory
(BCFT)\cite{cardy-84,cardy-05}.

The aim of this paper is twofold: on the one hand we give a
detailed description of the methods and the results reported
briefly in Ref. \cite{cc-06}, and on the other we generalize these
results and give some new physical insights. Several results, in
fact, appear here for the first time. The paper is organized as
follows. In Sec.~\ref{secpath} we introduce the path-integral
formalism that is applied in the following section \ref{secCFT} to
one-dimensional critical systems by means of CFT. In
Sec.~\ref{secint} we consider two simple 1D models whose
non-equilibrium dynamics is exactly solvable to check the
correctness of the previous results and to extend our findings to
gapped and lattice systems. In Sec.~\ref{HD} we generalize the
method to higher dimensions. In the last section \ref{phint} we
analyze all our results and explain most of the general findings
in quantum quenches by means of simple physical arguments. We
broadly discuss some open questions and possible future
developments.

\section{Path integral formulation and surface criticality}
\label{secpath}

Let us consider a lattice quantum theory in $d$ space dimensions.
The lattice spacing is $a$, and the lattice variables are labelled
by a discrete vector variable ${\bf r}$. Time is considered to be
continuous. The dynamics of the theory is described by the
hamiltonian $H$. Suppose we prepare this system in a state
$|\psi_0\rangle$ that is not an eigenstate of $H$ and unitarily
evolve it according to $H$. The expectation value of a local
operator ${\cal O}(\{ {\bf r}_i\})$ at time $t$ is \be \langle
{\cal O}(t,\{ {\bf r}_i\})\rangle= \langle \psi_0 | e^{i H t}
{\cal O}(\{ {\bf r}_i\})e^{-i H t}|\psi_0 \rangle\,. \ee
We modify this time-dependent expectation value as \be \langle
{\cal O}(t,\{ {\bf r}_i\})\rangle= Z^{-1} \langle \psi_0 | e^{i H
t-\e H} {\cal O}(\{ {\bf r}_i\}) e^{-i H t-\e H}| \psi_0
\rangle\,, \label{Oexp} \ee where we have included damping factors
$e^{-\e H}$ in such a way as to make the path integral
representation of the expectation value absolutely convergent. The
normalization factor $Z=\langle\psi_0|e^{-2\e H}|\psi_0 \rangle$
ensures that the expectation value of the identity is one. At the
end of the calculation we shall set $\e$ to zero.

Eq. (\ref{Oexp}) may be represented by an analytically continued
path integral in imaginary time over the field variables
$\phi(\tau,{\bf r})$, with initial and final values weighted by
the matrix elements with $|\psi_0\rangle$:
\begin{equation}
\int[d\phi(\tau,{\bf r})]\langle\psi_0|\phi(\tau_2,{\bf r})\rangle
\langle\phi(\tau_1,{\bf
r})|\psi_0\rangle\,e^{-\int_{\tau_1}^{\tau_2}L[\phi]d\tau}
\end{equation}
where $\int_{\tau_1}^{\tau_2}L[\phi]d\tau$ is the (euclidean)
action. The operator ${\cal O}$ is inserted at $\tau=0$, and the
width of the slab is $2\e$. $\tau_1$ and $\tau_2$ should be
considered as real numbers during the calculation, and only at the
end should they be continued to their effective values $\pm\e-i
t$. In this way we have reduced the real-time non-equilibrium
evolution of a $d$ dimensional systems to the thermodynamics of a
$d+1$ field theory in a slab geometry with the initial state
$|\psi_0\rangle$ playing the role of boundary condition at both
the borders of the slab. The validity of our results relies on the
technical assumption that the leading asymptotic behavior given by
field theory, which applies to the Euclidean region (large
imaginary times), may simply be analytically continued to find the
behavior at large real time.

Eq. (\ref{Oexp}) can be in principle used to describe the time
evolution in any theory and for all possible initial conditions.
Unfortunately, in confined geometries only a few field theories
with very specific boundary conditions can be solved analytically
in such a way to have results that can be continued from real to
complex values. Some examples of these will be given in Sec.
\ref{HD}. (An interesting case concerns integrable massive
boundary field theories as for example the recent application to
the dynamics of coupled 1D condensates in Ref. \cite{gdlp-07}.)
However, the treatment greatly simplifies if a system is at or
close to a (quantum) phase transition. In fact, in this case, we
can use the powerful tools of Renormalization Group (RG) theory of
boundary critical phenomena (see, e.g. \cite{dd}). For example, in
the case of a scalar order parameter (corresponding to the Ising
universality class), the slab geometry of above is usually
described by the action \cite{dd} \be S(\phi)=\int d^d r
\int_{\tau_1}^{\tau_2} d\tau \left( \frac{1}{2}(\partial
\phi)^2+\frac{m^2}{2} \phi^2+\frac{g}{4!}\phi^4\right)+ \int d^d r
\,{c}[\phi(\tau=\tau_1,r)+\phi(\tau=\tau_2,r)]\,, \label{LGH} \ee
where $m^2$ measures the distance from criticality, $g>0$ ensures
the stability in the broken-symmetry phase, and $c$ denotes the
surface enhancement, i.e. the difference of the interactions on
the surface with respect to the bulk. The value of $c$ depends on
the boundary conditions. For example, the choice $c=+\infty$
forces the field to vanish on the boundary, thus corresponding to
Dirichlet boundary conditions. On the other hand $c=-\infty$
forces the field to diverge at the boundary. Finite values of $c$
correspond to intermediate boundary conditions. According to the
RG theory, the different boundary universality classes are
characterized by the fixed point of $c$. A complete RG analysis
(see, e.g.,\cite{dd}) shows that the possible fixed points of $c$
are $c^*=\infty,0,-\infty$, with $c^*=\pm\infty$ stable and
$c^*=0$ unstable. Any other value of $c$ flows under RG
transformations to $\pm \infty$. These universality classes are
called ordinary ($c^*=\infty$), special ($c^*=0$), and
extraordinary ($c^*=-\infty$) \cite{dd}.

\begin{figure}[b]
\centering
\includegraphics[width=16cm]{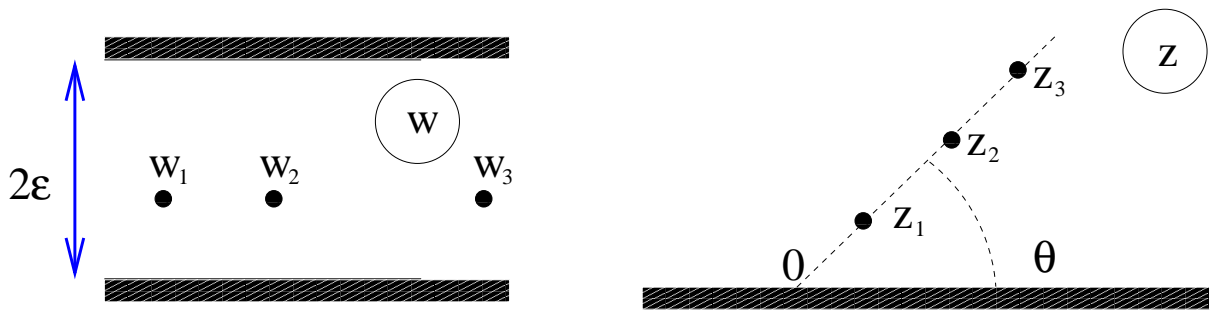}
\caption{Left: Space-imaginary time region in (\ref{Oexp}).
${\rm Im} w_i= \tau$, that will be analytically continued to $\tau\to \t0+it$.
Right: Conformal mapping of the left geometry to the upper half-plane
(c.f. Eq. (\ref{logmap})). Note that ${\rm arg} z_i= \theta=\pi \tau/2\t0$.}
\label{figpiO}
\end{figure}

Thus, for the purpose of extracting the asymptotic behavior, as
long as  $|\psi_0\rangle$ is translationally invariant, we may
replace it by the appropriate RG-invariant boundary state
$|\psi_0^*\rangle$ to which it flows. The difference may be taken
into account, to leading order, by assuming that the RG-invariant
boundary conditions are not imposed at $\tau=\tau_1$ and $\tau_2$
but at $\tau=\tau_1-\tau_0$ and $\tau=\tau_2+\tau_0$. In the
language of boundary critical behavior, $\tau_0$ is called the
extrapolation length \cite{dd}. It characterizes the RG distance
of the actual boundary state from the RG-invariant one. It is
always necessary because scale-invariant boundary states are not
in fact normalizable\cite{cardy-05}. It is expected to be of the
order of the typical time-scale of the dynamics near the ground
state of $H_0$, that is the inverse gap $m_0^{-1}$. (This can be
checked explicitly for a free field theory, see later). The effect
of introducing $\tau_0$ is simply to replace $\e$ by $\e+\tau_0$.
The limit $\e\to0+$ can now safely be taken, so the width of the
slab is then taken to be $2\tau_0$. For simplicity in the
calculations, in the following we will consider the equivalent
slab geometry between $\tau=0$ and $\tau =2\tau_0$ with the
operator ${\cal O}$ inserted at $\tau=\t0+it$. This is illustrated
in the left part of Fig.~\ref{figpiO}. In fact, in this geometry
we can consider products of operators at different times $t_j$, by
analytically continuing their labels to $\tau_0+it_j$.

Let us discuss to what initial conditions the fixed points
correspond. In the case of free boundary conditions, when the
order parameter is unconstrained, the continuum limit forces it to
vanish there, so it corresponds to the ordinary transition
($c=\infty$). For a lattice model, this corresponds to a
completely disordered initial state (e.g. for the Ising chain in a
transverse field corresponds to infinite field). In contrast for
non-vanishing fixed boundary conditions (e.g. + or $-$ for
Ising-like systems), the continuum limit makes the order parameter
to diverge at the boundary, thus corresponding to the
extraordinary transition $(c=-\infty)$.

\section{One space dimension and Conformal Field Theory}
\label{secCFT}

In this section we specialize the methods just introduced to the case when
$H$ is at a quantum critical point whose long-distance behavior is given by a
1+1-dimensional CFT, with dispersion relation $\Omega_k=v|k|$.
We set $v=1$ in the following. RG-invariant boundary conditions then correspond
to conformally invariant boundary states.
In this case the correlation functions are accessible through the powerful
tools of boundary CFT.

The main property we will repeatedly use in the following is the relation
of correlation functions of (primary) operators among two geometries connected
by a conformal transformation.
For example, the slab geometry of above is just a two-dimensional
strip whose points are labelled by a complex number $w=r+i\tau$
with $0<{\rm Im} w<2\tau_0$.
The strip can be obtained from the upper half-plane (UHP) ${\rm Im} z>0$
by the conformal mapping
\be
w(z)=\frac{2\t0}{\pi}\log z\,,
\label{logmap}
\ee
with the images of points at the same imaginary time on the strip
lying along ${\rm arg} z_i= \theta=\pi \tau/2\t0$.
In the case where $\cal O$ is a product of local \em primary \em scalar
operators $\Phi_i(w_i)$, the expectation value in the strip is related to the
one in the UHP by the standard transformation
\be
\langle\prod_i\Phi_i(w_i)\rangle_{\rm strip}
=\prod_i|w'(z_i)|^{-x_i} \langle\prod_i\Phi_i(z_i)\rangle_{\rm UHP}\,,
\ee
where $x_i$ is the bulk scaling dimension of $\Phi_i$. Note that
the (eventual) expectation values of the $\Phi_i$ in the ground state of $H$
are supposed to have been subtracted off.
The asymptotic real time dependence is obtained via the analytic continuation
$\tau\to\t0+it$, and taking the limit $t,r_{ij}\gg \t0$.

In the following subsections we apply these methods to some specific cases.

\subsection{The one-point function}

In the UHP, the one-point function of a scalar primary field with bulk scaling
dimension $x$ is
$\langle\Phi(z)\rangle_{\rm UHP}= A_b^\Phi [2{\rm Im}(z)]^{-x}$, as a
simple consequence of scaling invariance.
The normalization factor $A^\Phi_b$ is a non-universal amplitude.
In CFT the normalizations are chosen in such a way that the bulk two-point
functions have  unit amplitude (i.e.
$\langle\Phi(z_1) \Phi(z_2)\rangle_{\rm bulk}=|z_2-z_1|^{-2x}$). This choice
fixes unambiguously the amplitude $A^\Phi_b$ that turns out to depend both on
the considered field $\Phi$ and on the boundary condition on the real axis
$b$.
It vanishes if $\Phi$ corresponds to an operator whose expectation value
in $|\psi_0\rangle$ vanishes, and thus $\langle\Phi(t)\rangle=0$, for
all times.

When the primary field is not vanishing on the boundary,
performing the conformal mapping (\ref{logmap}) we obtain
\be
\langle \Phi(w)\rangle_{\rm strip}= |w'(z)|^{-x} \langle
\Phi(z(w))\rangle_{\rm UHP}= A^\Phi_b \left[\frac{\pi}{4\t0} \frac{1}{\sin(\pi
\tau/(2\t0))}\right]^{x}
\label{phistrip}
\ee
that continued to real time $\tau=\t0+i t$ gives
\be
\langle\Phi(t)\rangle=
A^\Phi_b \left[\frac{\pi}{4\t0} \frac{1}{\cosh(\pi t/(2\t0))}\right]^{x}
\simeq A^\Phi_b \left(\frac{\pi}{2\t0}\right)^{x} e^{-x\pi t/2\t0}\,.
\label{onepoint}
\ee
Thus the order parameter (and any other observables described by a
primary field) decays exponentially in time to the ground-state value,
with a non-universal
relaxation time $t_{\rm rel}^{\cal O}=2\t0/x_{\cal O} \pi$.
The ratio of the relaxation times of two different observables equals
the inverse of the ratio of their scaling dimensions and it is
then {\it universal}.

The normalization factor $A^\Phi_b$ is known for the simplest boundary
universality classes \cite{cl-91}. In the case of $\Phi$ being the order
parameter and the boundary condition is fixed ($\psi_0(x)=\infty$)
$A^\Phi_b$ is $1$ for the free boson and $2^{1/4}$ for the Ising model.

An important exception to this law is the local energy density (or
any piece thereof). This corresponds to the $tt$ component of the
energy-momentum tensor $T_{\mu\nu}$. In CFT this is not a primary
operator. Indeed, if it is normalized so that $\langle
T_{\mu\nu}\rangle_{\rm UHP}=0$, in the strip \cite{cardybn-86}
$\langle T_{tt}({\bf r},\tau)\rangle=\pi c/24(2\tau_0)^2$ (where $c$ is
the central charge  of the CFT) so that it does not decay in time.
Of course this is to be expected since the dynamics conserves
energy. A similar feature is expected to hold for other local
densities corresponding to globally conserved quantities which
commute with $H$, for example the total spin in isotropic models.

\subsection{The two-point function}

In the case of the one-point function, scaling invariance was
enough to fix the functional dependence on the position in the
UHP. However, the form of the two-point function depends
explicitly on the boundary universality class and on the operator
considered. In the following subsections we will consider the
equal-time correlation function for the order parameter in the
gaussian and in the Ising universality classes that are easily
treated in full generality. At the quantum level they describe
(among the other things) a chain of harmonic oscillators
(explicitly considered in Sec. \ref{secharm}) and the Ising model
in a transverse field (whose real time evolution has been
considered in Refs. \cite{bm1,bm2,ir-00,sps-04} and is briefly
reviewed in Sec. \ref{secIs}). Finally we will discuss the general
form of the two-point function for asymptotically large time and
distance, that can be obtained from general CFT arguments.

\subsubsection{The gaussian model}

The content of this subsection has been already reported in Ref.
\cite{cc-05}, during the study of the time evolution of the
entanglement entropy, that transforms like the two-point function
of a primary field in a boundary gaussian theory \cite{cc-04}. We
report it here for sake of completeness.

For a free boson the two-point function in the UHP is
\cite{cardy-84} \be \langle\Phi(z_1) \Phi(z_2) \rangle_{\rm
UHP}=\left
(\frac{z_{1\bar2}z_{2\bar1}}{z_{12}z_{\bar1\bar2}z_{1\bar1}z_{2\bar2}}\right)^x\,,
\label{bosUHP} \ee with $z_{ij}=|z_i-z_j|$ and $z_{\bar
k}={\overline{z_k}}$. Note that $\Phi$ is not the gaussian field
$\theta(z)$, but its exponential $\Phi(z)=e^{i\theta(z)}$ (see
Sec. \ref{secharm}).

Under the conformal mapping (\ref{logmap}) we obtain the two-point function
on the strip at imaginary time $\tau$, at distance $r$ apart
\bea
\langle\Phi(r,\tau) \Phi(0,\tau) \rangle_{\rm strip}&=&
|w'(z_1)|^{-x} |w'(z_2)|^{-x} \langle
\Phi(z_1(w))\langle \Phi(z_2(w))\rangle_{\rm UHP}= \nonumber \\ &=&
\left[\left(\frac{\pi}{2\t0}\right)^2 \frac{\cosh (\pi r/2\t0)- \cos
(\pi \tau/\t0)}{8 \sinh^2 (\pi r/4\t0) \sin^2(\pi
\tau/2\t0)}\right]^x,
\label{2ptstrip}
\eea
that continued to real time $\tau=\t0+i t$ gives
\begin{equation}
\langle\Phi(r,t) \Phi(0,t) \rangle=
\left[\left(\frac{\pi}{2\t0}\right)^2
\frac{\cosh(\pi r/2\t0)+\cosh(\pi t/\t0)}
{8 \sinh^2(\pi r/4\t0) \cosh^2(\pi t/2\t0)}\right]^x\,.
\end{equation}
In the case where $r/\t0$ and $t/\t0$ are large this simplifies to
\be
\langle\Phi(r,t) \Phi(0,t) \rangle=
(\pi/2\t0)^{2x}\left(\frac{e^{\pi r/2\t0} +e^{\pi
t/\t0}} {e^{\pi r/2\t0}\cdot e^{\pi t/\t0}}\right)^{x}\propto
\begin{cases}
e^{-x \pi t/\t0}  \qquad &{\rm for}\; t<r/2\\
e^{-x \pi r/2\t0} \qquad &{\rm for}\; t>r/2\\
\end{cases}\,.
\label{freeboson}
\ee
i.e. the two point function at fixed $r$ decays exponentially in time up to
$t^*=r/2$ and then saturates to a value that depends exponentially on the
separation.

In the case of fixed initial condition, with one-point
function given by Eq. (\ref{onepoint}), the connected correlation function is
\be
\langle\Phi(r,t) \Phi(0,t) \rangle_{\rm conn}=
\langle\Phi(r,t) \Phi(0,t) \rangle- \langle \Phi(0,t)\rangle^2\propto
\begin{cases}
0 & {\rm for}\; t<r/2\,,\\
e^{-x\pi r/2\t0}- e^{-x\pi t/\t0} & {\rm for}\; t>r/2\,,
\end{cases}
\ee
i.e. correlations start developing at $t^*=r/2$ and, being $t\gg \t0$, at
$t^*$ the connected two-point function almost immediately jumps to its
asymptotic value.
In the case of disordered initial conditions ($\psi_0(r)=0$), connected and
full correlation functions are equal.


\subsubsection{The Ising universality class}

For the Ising model the two-point function in the UHP is \cite{cardy-84}
\be
\langle\Phi(z_1) \Phi(z_2) \rangle_{\rm UHP}=\left
(\frac{z_{1\bar2}z_{2\bar1}}{z_{12}z_{\bar1\bar2}z_{1\bar1}z_{2\bar2}}\right)^{1/8}F(\eta)\,,
\ee
where $F(\eta)$ is given by
\be
F(\eta)=\frac{\sqrt{1+\eta^{1/2}}\pm
\sqrt{1-\eta^{1/2}}}{\sqrt{2}}\,,
\label{defF}
\ee
and $\eta$ is the four-point ratio
\be
\eta= \frac{z_{1\bar1}z_{2\bar2}}{z_{1\bar 2}z_{2\bar1}}\,.
\ee
The sign $\pm$ depends on the boundary conditions. $+$
corresponds to fixed boundary conditions and and $-$
to disordered ones.

The only difference with respect to the gaussian case is that we
have also to map  $F(\eta)$ according to the conformal
transformation (\ref{logmap}). After simple algebra we have \be
\eta=\frac{2 \sin^2 (\pi\tau/2\t0)}{\cosh(\pi r/2\t0)-\cos(\pi
\tau/\t0)}\,, \ee and so \bea &&\langle\Phi(r,\tau) \Phi(0,\tau)
\rangle_{\rm strip}= \left[\left(\frac{\pi}{2\t0}\right)^2
\frac{\cosh (\pi r/2\t0)- \cos (\pi \tau/\t0)}{8
\sinh^2 (\pi r/4\t0) \sin^2(\pi \tau/2\t0)}\right]^{1/8}\frac{1}{\sqrt2}\nonumber\\
&&\times \left[\sqrt{1+ \frac{\sqrt2 \sin (\pi
\tau/2\t0)}{\sqrt{\cosh(\pi r/2\t0)-\cos(\pi \tau/\t0)}}}\pm \sqrt{1-
\frac{\sqrt2 \sin (\pi \tau/2\t0)}{\sqrt{\cosh(\pi r/2\t0)-\cos(\pi
\tau/\t0)}}} \right]\,.
\eea
Analytically continuing to real time $\tau=\t0+i t$ we obtain
\bea
&&\langle\Phi(r,t) \Phi(0,t) \rangle=
\left[\left(\frac{\pi}{2\t0}\right)^2
\frac{\cosh(\pi r/2\t0)+\cosh(\pi t/\t0)}
{8 \sinh^2(\pi r/4\t0) \cosh^2(\pi t/2\t0)}\right]^{1/8}\frac{1}{\sqrt2}\nonumber\\
&&\times \left[\sqrt{1+ \frac{\sqrt2 \cosh(\pi
t/2\t0)}{\sqrt{\cosh(\pi r/2\t0)+\cosh(\pi t/\t0)}}}\pm \sqrt{1-
\frac{\sqrt2 \cosh(\pi t/2\t0)}{\sqrt{\cosh(\pi r/2\t0)+\cosh(\pi
t/\t0)}}} \right]\,,
\eea
that for $r/\t0$ and $t/\t0$ much larger than $1$ simplifies to
\be
(\pi/2\t0)^{1/4}\frac{1}{\sqrt2}\left(\frac{e^{\pi r/2\t0} +e^{\pi
t/\t0}} {e^{\pi r/2\t0}\cdot e^{\pi t/\t0}}\right)^{1/8} \left[\sqrt{1+
\frac{ e^{\pi t/2\t0}}{\sqrt{e^{\pi r/2\t0}+e^{\pi t/\t0}}}}\pm
\sqrt{1- \frac{ e^{\pi t/2\t0}}{\sqrt{e^{\pi r/2\t0}+e^{\pi t/\t0}}}}
\right]\,.
\ee

Note that the exponential terms in the square root are always $\ll 1$.
Thus for fixed boundary condition we get the free boson result
Eq. (\ref{freeboson}) with $x=1/8$.
For the connected part we need to subtract $\langle \Phi(0,t)\rangle^2$
given by Eq. (\ref{onepoint}) with $A^\Phi_+=2^{1/4}$.
We finally obtain
\be
\langle\Phi(r,t) \Phi(0,t) \rangle_{\rm conn}=
\langle\Phi(r,t) \Phi(0,t) \rangle- \langle \Phi(0,t)\rangle^2\propto
\begin{cases}
0 & {\rm for}\; t<r/2\,,\\
e^{-\pi r/16\t0}- e^{- \pi t/8\t0} & {\rm for}\; t>r/2\,.
\end{cases}
\ee
Thus, also for the Ising model with fixed boundary conditions, connected
correlations start developing at $t=t^*=r/2$.

In the case of disordered initial condition, we have
\be
\langle\Phi(r,t) \Phi(0,t) \rangle
\propto \left(\frac{e^{\pi r/2\t0} +e^{\pi
t/\t0}} {e^{\pi r/2\t0}\cdot e^{\pi t/\t0}}\right)^{1/8}
 \frac{ e^{\pi t/2\t0}}{\sqrt{e^{\pi r/2\t0}+e^{\pi t/\t0}}}\sim
\begin{cases}
  e^{-\pi (r-3/2 t)/4\t0 } &\quad {\rm for}\, t<r/2\,, \\
  e^{-\pi r /16\t0 }   &\quad {\rm for}\, t>r/2\,,
\end{cases}
\ee
resulting in an exponential space dependence even for $t<r/2$ (clearly
in this case the connected correlation function equals the full one).

\subsubsection{The general two point-function}

From the results reported for the gaussian and Ising models, it is
now relatively simple to understand the general properties of the
time dependence of the two-point function in the very general
case. The two-point function in the half-plane has the general
form \cite{cardy-84} \be \langle\Phi(z_1) \Phi(z_2) \rangle_{\rm
UHP}=\left
(\frac{z_{1\bar2}z_{2\bar1}}{z_{12}z_{\bar1\bar2}z_{1\bar1}z_{2\bar2}}\right)^x
F(\eta)\,, \label{2ptgen} \ee where the function $F(\eta)$ depends
explicitly on the considered model. Under the conformal map to the
strip we know that the first part of Eq. (\ref{2ptgen}) transforms
according to Eq. (\ref{2ptstrip}). Thus we need only to map
$F(\eta)$ that, in the general case, is an unknown function.
However, during the study of the Ising model we showed that the
analytical continuation of $\eta$ for $t,r\gg\t0$ is \be \eta \sim
\frac{e^{\pi t/\t0}}{e^{\pi r/2\t0}+ e^{\pi t/\t0}}\,. \ee Thus
for $t<r/2$ we have $\eta\sim e^{\pi (t-r/2)/\t0}\ll 1$ and in the
opposite case $t>r/2$ we have $\eta\sim1$. As a consequence to
have the asymptotic behavior of the two-point function we only
need to know the behavior close to $\eta\sim0$ (i.e. the behavior
close to the surface) and for $\eta\sim 1$ (i.e. deep in the
bulk). Fortunately they are both exactly known. Indeed when
$\eta\sim1$ the two points are deep in the bulk, meaning $F(1)=1$.
Instead for $\eta\ll 1$, from the short-distance expansion, we
have \be F(\eta)\simeq (A^\Phi_b)^2\eta^{x_b}, \ee where $x_b$ is
the boundary scaling dimension of the leading boundary operator to
which $\Phi$ couples and $A^\Phi_b$ is the bulk-boundary operator
product expansion coefficient that equals the one introduced in
Eq. (\ref{onepoint}) [see e.g. Ref. \cite{cl-91}].

All the previous observations and the explicit calculations of the previous
sections lead for $t>r/2$ to
\be
\langle\Phi(r,t)\Phi(0,t)\rangle \propto e^{-x\pi r/2\t0}\,,
\ee
while for $t<r/2$ we get
\be
\langle\Phi(r,t)\Phi(0,t)\rangle \propto (A^\Phi_b)^2 e^{-x\pi t/\t0}
 \times  e^{\pi x_b(t-r/2)/\t0}\,.
\ee
Note that if $\langle\Phi\rangle\neq0$, $x_b=0$ and the last factor is absent.
The leading term is then just $\langle\Phi\rangle^2$.
Thus the leading term in the connected two-point function vanishes for $t<r/2$,
and its first non-vanishing contributions is given by subleading terms either
in $F$ or in the bulk-boundary short-distance expansion.

It is very interesting that we only have to know the behavior as
$\eta\to0$ and $1$ to get the results we need for large $r$ and
$t$. However, we stress that only a complete calculation (as those
performed in the preceding sections) gives the full analytic
structure of the CFT result needed to justify the analytical
continuation from imaginary to real time. Moreover, the behavior
within a distance $O(\tau_0)$ of the horizon $r=2t$ depends on the
detailed form of $F$.

\subsection{Correlations functions at different times}

\begin{figure}[t]
\centerline{\epsfig{width=14cm,file=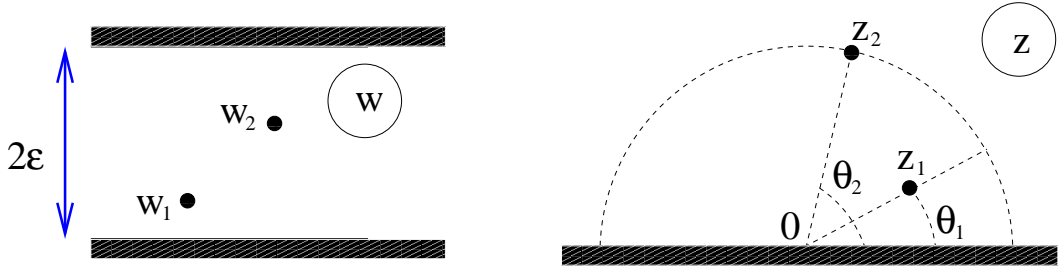}}
\caption{
Left: Space-time region for the correlation functions at different times.
Right: Conformal mapping to the upper half-plane.}
\label{fig2}
\end{figure}

Let us consider the case of the two-point function calculated at
different real times $\langle\Phi(r,t) \Phi(0,s)\rangle$.
This is again obtained by mapping the imaginary time strip to the UHP, but in
this case the two points are $w_1=r+i\tau_1$ and $w_2=0+i\tau_2$,
that, at the end of the calculation, must be analytically continued
to $\tau_1=\t0+i t$ and $\tau_2=\t0+i s$. See Fig. \ref{fig2} for a
pictorial representation of the space-time domain in the strip and the
resulting mapping to the UHP.

Let us start the discussion with the free boson.
The distances appearing in Eq. (\ref{bosUHP}) are
\bea
z_{12}^2&=&1+e^{\pi r/\t0}-2e^{\pi r/2\t0} \cos (\theta_1-\theta_2)\,,\qquad
z_{1\bar1}=2\sin\theta_1\,,\nonumber\\
z_{1\bar2}^2&=&1+e^{\pi r/\t0}-2e^{\pi r/2\t0} \cos (\theta_1+\theta_2)\,,
 \qquad z_{2\bar2}=2 e^{\pi r/2\t0}\sin\theta_2\,,
\eea where $\theta_i=\pi \tau_i/2\t0$. Thus the correlation
function on the strip is \bea &&\langle\Phi(r,\tau_1)
\Phi(0,\tau_2) \rangle_{\rm strip}= |w'(z_1)|^{-x} |w'(z_2)|^{-x}
\langle \Phi(z_1(w))\langle \Phi(z_2(w))\rangle_{\rm UHP}=
\\ &&=
\left[\left(\frac{\pi}{2\t0}\right)^2
\frac{\cosh (\pi r/2\t0)- \cos (\pi (\tau_1+\tau_2)/2\t0)}{4
\sin(\pi \tau_1/2\t0)\sin(\pi \tau_2/2\t0)
(\cosh (\pi r/2\t0)- \cos (\pi (\tau_1-\tau_2)/2\t0))
}\right]^{x}\nonumber ,
\eea
that for $\tau_1=\tau_2$ reduces to Eq. (\ref{2ptstrip}) as it should.
Continuing to real times and considering $r,t,s,|t-s| \gg\t0$ we obtain
\be
\displaystyle
\left(\frac{\pi}{2\t0}\right)^{2\Delta}
\frac{e^{\frac{\Delta \pi r}{\t0}} +e^{\frac{\Delta \pi}{\t0} (t+s)}}{
e^{\frac{\Delta \pi}{\t0} (t+s)}(e^{\frac{\Delta \pi r}{\t0}} +e^{\frac{\Delta \pi}{\t0} |t-s|})} =
\begin{cases}
e^{-x\pi(t+s)/4\t0}\quad &{\rm for}\, r>t+s\,,\\
e^{-x\pi r /4\t0}\quad &{\rm for}\, t-s<r<t+s\,,\\
e^{-x\pi|t-s|/4\t0}\quad &{\rm for}\, r<|t-s|\,.
\end{cases}
\ee

Following the line sketched in the previous subsection, it is easy to
generalize this result to the most general CFT.
In the case of a theory with fixed initial conditions (i.e.,
$\langle\Phi\rangle\neq0$) the asymptotic result is the same as before,
with only the crossover points being affected by the precise
expression for $F(\eta)$.
Instead, in the case where $\langle\Phi\rangle=0$, the first
case gains an additional factor $e^{-\pi x_b(t+s-r)/4\tau_0}$.

Note that the autocorrelation function (i.e. $r=0$) has only an exponential
dependence on the time separation $t-s$ and does not exhibit aging in this
regime.

\subsection{Evolution with boundaries}
\label{evbou}

\begin{figure}[t]
\centerline{\epsfig{width=14cm,file=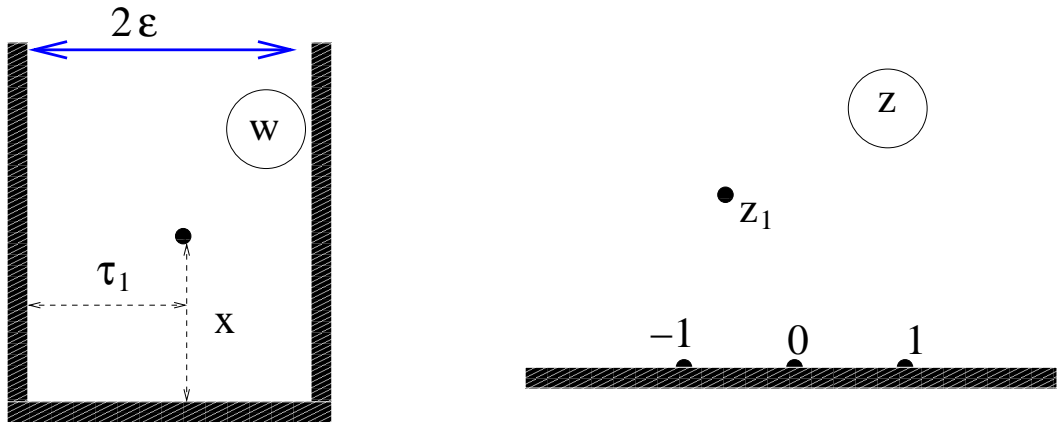}}
\caption{Left: Space-time region for the one-point function in a
boundary (at $r=0$) geometry.
Note $w=\tau+ir$. Right: Conformal mapping to the upper-half plane,
c.f. Eq. (\ref{mapsin}).}
\label{fig3}
\end{figure}

We now consider the case of time evolution of a half-chain with some boundary
condition at $r=0$. For simplicity we assume that the (conformal) boundary
condition is of the same kind of the initial boundary condition
(for example we fix all the spins at $t=0$ and at the boundary $r=0$ to point
in the same direction).
The space-time region  we have to consider is depicted in Fig. \ref{fig3}.
If different initial and boundary were considered, one needs to insert
boundary conditions changing operators at the corners of the figure.

The $w$ plane is mapped into the UHP by
\be
z(w)=\sin \frac{\pi w}{2\t0}\,,
\label{mapsin}
\ee
with the corners at $\pm \t0$ mapped to $\pm1$.
The mapping of $w_1$ is
\be
z_1\equiv z(w_1)= z(-\t0+\tau_1+i r)= -\cos (\pi \tau_1/2\t0)\cosh (\pi r/2\t0)
+i \sin (\pi \tau_1/2\t0)\sinh (\pi r/2\t0)\,.
\ee
In the $z$ plane the 1-point function is
\be
\langle \Phi (z_1)\rangle_{\rm UHP}\propto |{\rm Im} {z_1}|^{-x}\to
\left[\sin (\pi \tau_1/2\t0)\sinh (\pi r/2\t0)  \right]^{-x}\,,
\ee
and
\be
|w'(z_1)|^2= \left(\frac{2\t0}{\pi}\right)^2 \frac{1}{|1-z^2|}\propto
\frac{1}{\cosh (\pi r/\t0) -\cos(\pi \tau_1/\t0) }\,.
\ee
Thus on the strip we have
\be
\langle \Phi (w_1)\rangle_{\rm strip}=
|w'(z_1)|^{-x} \langle \Phi (w(z_1))\rangle_{\rm UHP}\propto
\left[ \frac{\sin^2 (\pi \tau_1/2\t0)\sinh^2(\pi r/2\t0)}{
\cosh (\pi r/\t0) -\cos(\pi \tau_1/\t0) }\right]^{-x/2}\,,
\ee
that continued to real time $\tau_1=i t$ is
\be
\langle \Phi (t,r)\rangle\propto
\left[
\frac{\cosh (\pi t/\t0)+\cosh (\pi r/\t0)}{
\cosh(\pi t/2\t0)^2\sinh^2 (\pi r/2\t0)}
\right]^{x/2},
\ee
and for $t,r\gg\t0$ simplifies to
\be
\langle \Phi (t,r) \propto \left[
\frac{e^{\pi r/\t0}+e^{\pi t/\t0}}{e^{\pi r/\t0}\cdot e^{\pi t/\t0}}
\right]^{x/2}=
\begin{cases}
e^{-\pi xt/2\tau_0}\quad &{\rm for}\, t<r\,,\\
e^{-\pi xr/2\tau_0}\quad &{\rm for}\, t>r\,.
\end{cases}
\ee
Note that in this case the characteristic time is $t^*=r$ and not $r/2$.
This explains also why the entanglement entropy of a semi-infinite chain with
free boundary condition at $x=0$, has characteristic time $t^*=r$ as firstly
noted in Ref. \cite{dmcf-06}.

\subsection{Discussion and interpretation of the CFT results}

All the correlation functions calculated so far display two very
general features: first there is a sharp horizon (or light-cone)
effect at $t=t^*=r/2$ (or $r$) resulting in a behavior before and
after $t^*$ completely different; second the asymptotic long-time
correlation functions are the same as those at finite temperature
$\beta_{\rm eff}=4\tau_0$. The light-cone effect is a very general
phenomenon and and will be discussed in section \ref{phint}. In
the following sections we will point out that also the ``effective
temperature'' is a general phenomenon, but it has some specific
CFT features that are worthy of comment.

In fact, it is very easy to understand the \em technical \em
reason why we find an effective temperature despite the fact that
we are studying a pure state at $T=0$. The finite temperature
correlations can be calculated by studying the field theory on a
cylinder of circumference $\beta=1/T$. In CFT a cylinder can be
obtained by mapping the complex plane with the logarithmic
transformation $\beta/(2\pi)\log z$. The form for the two-point
function in the slab depends in general on the function $F(\eta)$
--cf. Eq. (\ref{defF})-- but when we analytically continue and
take the limit of large real time, we find that effectively the
points
 are far from the boundary, i.e. at $\eta=1$.
Thus we get the same result as we would get if we conformally transformed
from the full plane to a cylinder, and from Eq. (\ref{logmap}) the effective
temperature is $\beta_{\rm eff}=4 \tau_0$.
A similar argument can be worked out for the multi-point functions as well.

An addtional comment concerns what we expect for correlation
functions of general operators, not only primary. We have clearly
seen that the local energy density does not relax, of course, and
this is consistent with its not being primary. But, we also know
that at finite temperature $\langle T_{tt}\rangle_\beta= \pi c/6
\beta^2$, that is perfectly compatible with the previous result
with $\beta_{\rm eff}=4\tau_0$. Furthermore, for large real times,
the two-point function of $T_{tt}-\langle T\rangle_{tt}$ does
behave as though it was at finite temperature. This means, in
particular, that the energy fluctuations in a large but finite
volume (the specific heat) are the same as those at finite
temperature. Thus one is tempted to extend this finite temperature
interpretation to non-primary operators, and indeed it is the
case. In fact, in the argument of above for the equivalence of the
long-time correlations and finite $T$, there are essentially three
steps. First, we need to write down the form of the correlation
function in the half-plane, but this only depends on special
conformal transformation and so is valid for any quasi-primary
operator like $T_{\mu\nu}$. Second we have to transform from the
slab to the half-plane: for non-primary operators this can have
some anomalous term. Finally we need to compare the large limit of
this to what one would get transforming directly from the cylinder
to the full plane. However, the two conformal transformations we
are comparing are both logarithmic (in one case $2\tau_0/\pi\log
z$, in the other case $\beta/2\pi\log z$), so the anomalous terms
should be the same. Thus the two correlations functions are the
same. This argument works for all quasi-primary operators. Then,
since we can get all the non-quasiprimaries by considering
successive operator product expansions with the stress tensor, it
also works for all operators.

\section{Exact real-time dynamics in simple integrable chains}
\label{secint}

The results of the previous section rely on the technical assumption that the
leading asymptotic behavior given by CFT, which applies to the euclidean
region (large imaginary times), may simply be analytically
continued to find the behavior at large real time. While such
procedures have been shown to give the correct behavior for the
time-dependent correlations in equilibrium, it is important to
check them in specific solvable cases for non-equilibrium evolution.

Thus, as a complement to the CFT calculations, in this section we consider
the \em real-time \em evolution of two simple analytically tractable models.
We solve the dynamics of a chain of coupled harmonic oscillators
and we review and re-analyze some known results for the Ising-XY chain in a
transverse magnetic field.
Beyond providing examples of the CFT results (with central charge $c=1$ and
$1/2$ respectively) in the critical case, these models allow us to take into
account the effects of a finite mass-gap and of the lattice.

\subsection{The chain of harmonic oscillators}
\label{secharm}

The simplest model with an exactly solvable non-equilibrium
dynamics is surely a chain of coupled harmonic oscillators with
hamiltonian \be
H=\frac12\sum_r\Big[\pi_r^2+m^2\phi_r^2+\sum_j\omega_j^2(\phi_{r+j}-\phi_r)^2
\Big]\,. \label{Hchaingen} \ee We introduce a coupling more
general than simple nearest-neighbor hopping so as to allow for a
general dispersion relation below. For simplicity we also assume
periodic boundary conditions. $\p_n$ and $\pi_n$ are the position
and the momentum operators of the $n$-th oscillator, with equal
time commuting relations \be [\p_m,\pi_n]=i\delta_{nm}\,\qquad
[\p_n,\p_m]=[\pi_n,\pi_m]=0\,. \ee

The hamiltonian can be written in diagonal form $H(m)=\sum_k
\Omega_k A^\dag_k A_k$ with modes \bea
A_k     &=&\frac{1}{\sqrt{2 \Omega_k}}(\Omega_k \p_k+i \pi_k)\,,\\
A_k^\dag&=&\frac{1}{\sqrt{2 \Omega_k}}(\Omega_k \p_{-k}-i \pi_{-k})\,,\\
\Omega_k^2&=&m^2+2\sum_j\omega_j\left(1-\cos(2\pi kj/N)\right)\,. \eea
Note that we use the same symbols for the operators and their
Fourier transforms ($\p_k=1/\sqrt{N} \sum_{n=0}^{N-1} e^{2\pi
ikn/N} \p_n$ and analogously for $\pi_k$).

We consider the scenario in which the system is prepared in a state
$|\psi_0\rangle$, that is  ground-state of $H(m_0)$, and at the time $t=0$ the
mass is quenched to a different value $m\neq m_0$.
We use the notation $\Omega_{0k}$ for the dispersion relation for $t<0$ and the
$\Omega_k$ for the one for $t>0$.

Since $\langle\psi_0|  \p_n |\psi_0\rangle=0$, the expectation
value of the field $\p_n$ vanishes at any time. This example in
fact corresponds to the quench from the disordered phase in the
language of the previous section. Thus we concentrate our
attention on the two-point function \be \langle \psi(t)| \p_n
\p_0|\psi(t) \rangle= \langle\psi_0|  \p^H_n(t) \p^H_0(t)
|\psi_0\rangle \,, \ee where we introduced the operator in the
Heisenberg picture $\p^H_n(t)$, whose time evolution is given by
\be \p^H_n(t)=\sum_{k=0}^{N-1} \sqrt{\frac{2}{N \Omega_k}}
\left(e^{i(p_k n-\Omega_k t)} A_k+e^{-i(p_k n-\Omega_k t)}
A_k^\dag\right)\,, \ee where $p_k=2\pi k/N$. Accordingly, the
product of the two fields is \be
\p^H_n(t)\p^H_0(t)=\frac{2}{N}\sum_{k,k'}\frac{1}{\sqrt{\Omega_k\Omega_{k'}}}
\left(e^{i(p_k n-\Omega_k t)} A_k+e^{-i(p_k n-\Omega_k t)} A_k^\dag\right)
\left(e^{-i\Omega_{k'} t} A_{k'}+e^{i\Omega_{k'} t} A_{k'}^\dag\right)\,.
\ee In order to have the time dependent two-point function we need
the expectation values of the bilinear combinations of $A$'s on
the initial state, that is annihilated by the $A_{0k}$. Thus it is
enough to write $A$'s as functions of $A_0$'s, i.e. \bea
\displaystyle A_k&=&\frac{1}{2}\left[ A_{0k}
\left(\sqrt{\frac{\Omega_k}{\Omega_{0k}}}+\sqrt{\frac{\Omega_{0k}}{\Omega_k}}\right)+
A_{0-k}^\dag\left(\sqrt{\frac{\Omega_k}{\Omega_{0k}}}-\sqrt{\frac{\Omega_{0k}}{\Omega_k}}\right)
\right]\equiv c_k A_{0k} +d_k A^\dag_{0-k}  \,,  \\
A^\dag_k&=&\frac{1}{2}\left[
A_{0k}^\dag\left(\sqrt{\frac{\Omega_k}{\Omega_{0k}}}+\sqrt{\frac{\Omega_{0k}}{\Omega_k}}\right)+
A_{0-k}
\left(\sqrt{\frac{\Omega_k}{\Omega_{0k}}}-\sqrt{\frac{\Omega_{0k}}{\Omega_k}}\right)
\right]\equiv c_k A^\dag_{0k} +d_k A_{0-k} \,, \eea leading to (we
understand $\langle\cdot \rangle=\langle
\psi_0|\cdot|\psi_0\rangle$) \bea \langle A_k A_{k'}\rangle &=&
\langle(c_k A_{0k} +d_k A^\dag_{0-k})(c_{k'} A_{0k'} +d_{k'}
A^\dag_{0-k'})\rangle=
c_k d_{k'} \langle A_{0k} A^\dag_{0-k'}\rangle= c_k d_k \delta_{k,-k'}\,, \\
\langle A_k A_{k'}^\dag\rangle      &=&
\langle(c_k A_{0k} +d_k A^\dag_{0-k})(c_{k'} A^\dag_{0k'} +d_{k'} A_{0-k'})\rangle=
c_k c_{k'} \langle A_{0k} A^\dag_{0k'}\rangle= c_k^2 \delta_{k,k'}\,,\\
\langle A_k^\dag A_{k'}\rangle      &=&
\langle(c_k A^\dag_{0k} +d_k A_{0-k})(c_{k'} A_{0k'} +d_{k'} A^\dag_{0-k'})\rangle=
d_k d_{k'} \langle A_{0-k} A^\dag_{0-k'}\rangle= d_k^2 \delta_{k,k'}\,,
\label{dk}\\
\langle A_k^\dag A_{k'}^\dag\rangle &=& \langle(c_k A^\dag_{0k}
+d_k A_{0-k})(c_{k'} A^\dag_{0k'} +d_{k'} A_{0-k'})\rangle= d_k
c_{k'} \langle A_{0-k} A^\dag_{0k'}\rangle= c_k d_k
\delta_{k,-k'}\,. \eea
Finally we arrive at \bea\displaystyle
\langle\p^H_n(t)\p^H_0(t)\rangle&=&\frac{2}{Na}\sum_k
\frac{1}{\Omega_k}\left[ c_k d_k e^{i(p_k n-2\Omega_k t)}+c_k^2 e^{ip_k
n}+d_k^2 e^{-i p_kn} +c_k d_k e^{-i(p_k n-2\Omega_k t)}\right]
\,,\label{finlat}
\eea
which, in the thermodynamic limit ($N\to\infty$), may be written
as \be \label{2ptgaussian} \langle\p_r^H(t)\p^H_0(t)\rangle-
\langle\p^H_r(0)\p^H_0(0)\rangle =\int_{\rm BZ} \frac{dk}{(2\pi)}
e^{ikr} \frac{(\Omega_{0k}^2-\Omega_k^2)(1-\cos(2\Omega_kt))}
{\Omega_k^2\Omega_{0k}}\,, \ee where the integral is on th first
Brillouin zone $|k|<\pi/a$. Note that for $t=0$ and for $m=m_0$
this two-point function reduces to the static one, as it should.
This result can also be found by integrating the Heisenberg
equations of motion for each mode. For future reference it is also
useful to write down explicitly the Fourier transform known as
momentum distribution function \be \rho(k)=
\frac{(\Omega_{0k}^2+\Omega_k^2)-(\Omega_{0k}^2-\Omega_k^2)
\cos(2\Omega_kt)}{\Omega_k^2\Omega_{0k}}\,. \label{MDF} \ee Note
that when considering correlation functions in momentum space, a
long-time limit does not exist and we need to take the time
average, in contrast to what happens in real space.

\subsubsection{The continuum limit}

In Eq. (\ref{finlat}) everything is completely general and applies
to any chain with finite lattice spacing. Let us know discuss the
continuum limit that is achieved by sending  $N\to\infty$ in such
a way that $(1/N) \sum_k \to \int_{-\infty}^{\infty} d p/(2\pi)$,
$p_n\to p$, and $\Omega_k^2 \to \Omega_p^2=m^2+p^2$.

In this limit the correlation function becomes ($\Delta m^2=m_0^2-m^2$)
\be
G(r,t)\equiv \langle\p_r^H(t)\p_0^H(t)\rangle=
\int_{-\infty}^{\infty} \frac{dp}{2\pi} e^{i p r}
\frac{-\Delta m^2 \cos(2 \sqrt{p^2+m^2} t)
+ m^2+m_0^2 +2 p^2}{{(m^2+p^2)\sqrt{m_0^2+p^2}}} \,.
\label{Gcont}
\ee
Since a closed form for such integral is quite difficult to write down in the
most general case, we will only consider some particular cases.


Let us first consider the conformal evolution $(m=0)$ from an initial state
with a very large mass $m_0\to\infty$.
This should reproduce the CFT result previous section for all the time $t$,
since the correlations of the initial state shrink to a point. We have
\be
G(r,t)=
m_0\int_{-\infty}^{\infty} \frac{dp}{2\pi} e^{i p r}
\frac{1-\cos(2 p t)}{p^2}=m_0
\begin{cases}
0 \;&{\rm for}\, t<r/2\,, \\\displaystyle
\frac{1}{2}(2t-r)\; &{\rm for}\, t>r/2\,.
\end{cases}
\label{Gconf} \ee To compare such result with the conformal result
given by Eq. (\ref{freeboson}), we have to keep in mind that the
primary field is not the gaussian one $\p^H(r,t)$, but its
imaginary exponential. Thus we need  $\langle e^{i q \p^H(r,t)}
e^{-i q \p^H(0,t)}\rangle$, with an arbitrary $q$. Despite of the
apparent complexity of such correlator, it is very simple to
obtain it using the standard property of gaussian integrals \be
\langle e^{i q\p^H(r,t)} e^{-iq \p^H(0,t)}\rangle=
e^{-q^2\langle(\p^H(r,t)-\p^H(0,t))^2\rangle/2}=
e^{q^2(G(r,t)-G(0,t))}\,, \ee leading to \be \langle e^{i
q\p^H(r,t)} e^{-i q\p^H(0,t)}\rangle=
\begin{cases}
e^{-q^2 m_0 t}\qquad & {\rm for } \;t<r/2\,,\\
e^{-q^2 m_0 r/2}\qquad & {\rm for } \;t>r/2\,,
\end{cases}
\ee
that is exactly the same of Eq. (\ref{freeboson}) with $x_\Phi\propto q^2$ and
$\t0\propto m_0^{-1}$, confirming that $\tau_0$ is just proportional to the
correlation in the initial state, as its interpretation in terms of the
extrapolation length suggests.


The case $m=0$ and $m_0$ finite corresponds to a conformal evolution from
a generic state with correlations proportional to $m_0^{-1}$.
Thus we expect the CFT result
Eq. (\ref{freeboson}) to be true for asymptotic large times and separations.

\begin{figure}[t]
\centering
\includegraphics[width=\textwidth]{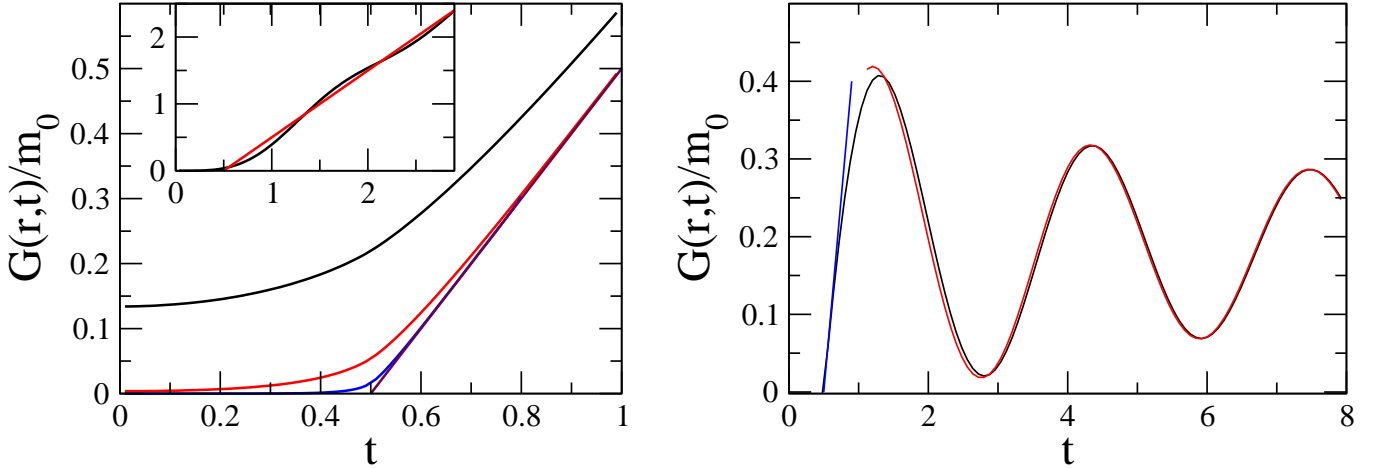}
\caption{Left: $G(r,t)/m_0$ given by Eq. (\ref{Gxt}) as function of $t$,
at fixed $r=1$.
Three different values of $m_0=10,3,1$ (from the bottom to the top) are shown.
Inset: Lattice effects showing the $\cos 4t$ oscillations on top of the
continuum  result.
Right: $G(r,t)/m_0$ given by the numerical integral of Eq. (\ref{Gmfin}) as
function of $t$, at fixed $m,r=1$. It is compared with the
asymptotic behavior for $0<2t-r\ll m^{-1}$ and for $t\gg r$.
}
\label{figGG}
\end{figure}

From Eq. (\ref{Gcont}), the two-point function of the gaussian
field is \be G(r,t)= \int_{-\infty}^{\infty} \frac{dp}{2\pi}
e^{ipr} \frac{-m_0^2 \cos(2 p t) +m_0^2 +2
p^2}{{p^2\sqrt{m_0^2+p^2}}}\,, \ee that can be written as \be
 2 \pi G(r,t)=2 K_0(m_0 r)+f(m_0r)-\frac{f(m_0(r-2t))+f(m_0(r+2t))}{2}\,,
\label{Gxt}
\ee
where $K_0(y)$ is the modified Bessel function and
\be
f(y)= 1+\frac{1}{2}G^{21}_{13}\left(\frac{ y^2}{4}\left|
\begin{matrix}
&  3/2 & \\
0 &1 &1/2
\end{matrix}\right.
\right)\,,
\ee
with $G^{21}_{13}$  the Meijer $G$-function (see e.g. \cite{math}).
Note that $f(x)$ is characterized by $f''(x)=-K_0(x)$ and $f(0)=f'(0)=0$.
In the limit $m_0\to\infty$, it is easy to show that
$G(r,t)$ reduces to the previous result.
In the left panel of Fig. \ref{figGG} ,
we report $G(r,t)$ as function of $t$ at fixed $r=1$ for
$m_0=10,3,1$. It is evident that a finite $m_0$ results in smoothing
the curve close to $t=r/2$ and giving an offset in zero.
Both the effects are more pronounced as $m_0$ decreases.
For large
$t$, independently on $m_0\neq 0$, we  have $G(r,t)= t+O(t^0)$, confirming
that the CFT result is correct for asymptotic large times.



The case with arbitrary $m$ and $m_0$ is quite cumbersome to be worked out
analytically and not really illuminating. For this reason we concentrate here
on the massive evolution from a state with $m_0\to\infty$.
In this case the correlation function reads
\be
G(r,t)=m_0
\int_{-\infty}^{\infty} \frac{dp}{2\pi} e^{ipr}
\frac{1-\cos(2 \sqrt{p^2+m^2} t)}{{(m^2+p^2)}} \,.
\label{Gmfin}
\ee
In the limit $t\to \infty$ the cosine term averages to zero, giving
\be
G(r,t=\infty)=m_0
\int_{-\infty}^{\infty} \frac{dp}{2\pi}
\frac{ e^{ipr}}{{(m^2+p^2)}} =\frac{m_0}{m} \frac{e^{-mr}}{2}\,.
\label{genmtinf}
\ee

To understand the time dependence, let us first note that, despite the
presence of a square-root, the integrand of Eq. (\ref{Gmfin}) is analytic in
$p$, since the square-root is the argument of the even cosine function. Thus we
can make the integral in the complex plane and use the Cauchy theorem. As long
as $r>2t$, the behavior for $p\to i\infty$ is dominated by $e^{i p r}$ and we
can safely close the contour path in the upper half-plane where the residue at
$p=i m$ is zero. As a consequence $G(r,t<r/2)=0$.

For $t>r/2$, the integral is more difficult, since we can not close the contour
in the upper half-plane, because the cosine is ``larger'' than $e^{ipr}$
for $p\to i\infty$. However, the approach to the asymptotic value is
easily worked out. In fact, for $t\gg r$, the term $e^{ipr}$ in the integral
(\ref{Gmfin}) can be approximated with $1$, since it is slowly
oscillating. Thus we have
\be
G(r,t\gg r)-G(r,t=\infty) \simeq -m_0
\int_{-\infty}^{\infty} \frac{dp}{2\pi}
\frac{\cos(2 \sqrt{p^2+m^2} t)}{{(m^2+p^2)}} = - m_0 f_m(t)\,,
\ee
where $f_m(t)=1/(2m)-t _1F_2(1/2; 1,3/2;-m^2t^2)$, that satisfies
$f'_m(t)=-J_0(2mt)$ and $f_m(0)=1/(2m)$. Note that $f_m(t)$ is
just the result for $r=0$.
In the complementary region $0<2t-r\ll m^{-1}$, the integral in
Eq. (\ref{Gmfin}) is dominated by the modes with $p\gg m$, and so it
can be described by the conformal result.

In the right panel of
Fig. \ref{figGG}, we plot the time-dependence of $G(r,t)$ at fixed $r$
as obtained by numerically integrating Eq. (\ref{Gmfin}) for $m=r=1$.
The plot shows that the conformal result $t-r/2$ describes the behavior
close to $t=r/2$, while for larger times the asymptotic expression
gives an excellent approximation.

We finally note that $G(r,t)$ at fixed $t$ displays spatial oscillations (for
$r<2t$, else it vanishes), as it can be simply realized by a stationary phase
argument.

\subsubsection{Lattice effects}
\label{lateff}

Another advantage of this simple model is that the effects of the lattice
can be easily understood.
In fact, by a stationary phase argument,
the dominant contribution to (\ref{2ptgaussian}) in the limits of
large $t$ and $r$ comes from where,
the group velocity $v_k\equiv\Omega'_k=r/2t$, independently of the explicit
form of $\Omega_k$.
Consequently, the two-point function of a gapless model
with dispersion relation $\Omega_k=2 \sin k/2$ differs from the continuum limit
previously derived for $t>r/2v_m$, where $v_m$ is the maximum group velocity.
In particular $G(r,t)$ receives a contribution from the slowest
mode ($k=\pi$ with $v_\pi=0$) whose effect is to add to Eq. (\ref{Gconf})
fast oscillations going as $\cos(2\Omega_\pi t)= \cos 4t$.
The resulting $G(r,t)$ is plotted in the inset of Fig. \ref{figGG} in the
gapless case.

Thus, lattice effects play a more important role in the cases where the
asymptotic result vanishes.
A typical example is the energy density that in this model is proportional
to $(\p_{r+1}-\p_r)^2$.
The continuum calculation would just give zero (the mean-energy is conserved),
but the approach to this value is governed by lattice effects and
is dominated by the smallest group velocity that
comes from the zone boundary at $|k|=\pi$.
In fact, just with a trivial calculation we have
\bea
\langle(\p_{r+1}-\p_r)^2\rangle&=&
2(\langle\p_0^2\rangle-\langle\p_0\p_1\rangle)=
\int\frac{dk}{2\pi} (1-\cos k) \frac{(1-\cos2\Omega_k t)}{2\sin^2k/2}=
\nonumber\\&=&
\int\frac{dk}{2\pi} (1-\cos2\Omega_k t)\propto J_0(4t)\sim t^{-1/2} \cos 4t\,,
\eea
that displays the typical $\cos 4t$ from the zone boundary at $|k|=\pi$.

The lattice dispersion relations are sensitive to the microscopical
details of the model and consequently quantities like the energy-density
show a dependence on this.
For example, a lattice massless fermion has dispersion relation
$\Omega_k=2\sin k/2$ 
and energy density given by $\sin^2 k$.
In this case the time evolution of the energy density is
\be
\int\frac{dk}{2\pi} \sin^2k \frac{(1-\cos2\Omega_k t)}{2\sin^2k/2}\,,
\ee
that for large times goes like $t^{-3/2}\cos 4t$, resulting in a different
power law compared to before.

If the system is quenched to a gapped $H$ with $\Omega_k^2=m^2+2(1-\cos k)$
the maximum group velocity corresponds to a non-zero wave number.
This gives rise to spatial oscillations in the correlation function (this is
true also in the continuum limit as already discussed in the previous
subsection).

\subsection{The Ising-XY chain in a transverse magnetic field}
\label{secIs}

The most studied one dimensional quantum spin model is the
so-called XY chain in a transverse field, defined by the
hamiltonian \be H=\sum_{j=1}^N \left[\frac{(1+\gamma)}{2}
\sigma^x_j \sigma_{j+1}^x +\frac{(1-\gamma)}{2}\sigma^y_j
\sigma_{j+1}^y- h\sigma_z\right]\,, \ee where $\sigma_i^{x,y,z}$
are the Pauli matrices, $\gamma$ is called anisotropy parameter,
and $h$ is the applied external transverse field. It is well known
that for any $\gamma\neq0$ the model undergoes a phase transition
at $h=1$, that is in the universality class of the Ising model
(defined by $\gamma=1$). For simplicity we will just consider the
Ising case, other values of $\gamma\neq 0$ being equivalent.

We consider the non-equilibrium unitary dynamics that follows from a
quench of the magnetic field at $t=0$ from $h_0$ to $h_1\neq h_0$.
Earlier works on this subject by Barouch et al. date back to
seventies \cite{bm1,bm2}.
Exploiting the mapping of this model onto a free fermion,
the time evolution of the transverse magnetization (that is {\it not} the
order parameter) was obtained exactly \cite{bm1}:
\be
m_z(t)=\frac{1}{\pi} \int_0^\pi dk \frac{(h_0-h_1)\sin^2k\cos(2\e_1 t)-
(\cos k-h_1)[(\cos k-h_0)(\cos k-h_1)+\sin^2 k]}{\e_0 \e_1^2}\,,
\ee
where $\e_i=\e(h_i)$ with $\e(h)=\sqrt{\sin^2 k+(h-\cos k)^2}$. In
Ref. \cite{bm1} it was shown that (for non-exceptional parameters) the
approach to the asymptotic value for $t\to\infty$ is of the form
$t^{-3/2}\cos 4t$. This is simply shown in the case of $h_0=\infty$
and $h_1=1$, when the integral simplifies and we obtain (see also \cite{ir-00})
\be
m_z(t)=\frac{1}{2}+\frac{1}{\pi} \int_0^\pi dk
\frac{\sin^2k\cos(2\e_1 t)}{4\sin^2 k/2}=
\frac{1}{2}+\frac{J_1(4t)}{4t}\,,
\ee
where $J_1$ is the Bessel function of the first kind whose
asymptotic expansion for large argument is
$J_1(x)\sim \sqrt{2/\pi x} \cos (x+\pi/4)$.

To understand this result we should keep in mind that
the transverse magnetization is {\it not} the order parameter:
it corresponds to the product of two ``disorder parameters''
at neighbor sites on the dual lattice. Thus it
must have the symmetry of an energy operator, for which CFT just
predicts a constant asymptotic result.
The power law term $t^{-3/2}\cos(4t)$ is just the lattice correction to the
asymptotic. This correction clearly shows the fermionic nature of the model
on the basis of what discussed in the previous section.

The asymptotic result of the two-point function for $t\to\infty$ has
been studied in Ref. \cite{sps-04}. Calling
$G_n=\langle\Phi(n,\infty) \Phi(0,\infty)\rangle$, in the case
$h_0=\infty$ it has been found
\be
G_n=\begin{cases}\displaystyle
\frac{1}{2^nh_1^n}\,,&{\rm for}\, h_1\geq1\,,\\
\frac{1}{2^n} \cos[n\arccos(h_1)] \,,&{\rm for}\, h_1\leq1\,,
\end{cases}
\ee
instead for $h_0=0$
\be
G_n=\begin{cases}
\frac{1}{2^n}\,,&{\rm for}\, h_1\geq1\,,\\
\frac{h_1^{n+1}}{2^n}\cosh\left[(n+1)\ln
\left(\frac{1+\sqrt{1-h_1^2}}{h_1} \right)\right]\,,& {\rm for} h_1\leq1\,,
\end{cases}
\ee
that for large $n$ decay exponentially with $n$.
It has been shown that $G_n$ is decaying exponentially with $n$
for general $h_0$ and $h_1$ \cite{sps-04}, although closed forms are not
available. The exponential decay is the prediction of CFT that (maybe
surprisingly) applies even far from the critical point $h_1=1$.

The time dependence of two-point function has been studied in
Ref. \cite{ir-00} by means of exact diagonalization of the model with open
boundary condition at the two ends $r=0,L$.
The results of interest for this paper are
\begin{itemize}
\item The connected two-point function
of $\sigma_z$ in the thermodynamic limit and at the critical point is
\be
\langle \s_r^z(t)\s_0^z(t)\rangle_c
=\left[\frac{r}{2t}J_{2r}(4t)\right]^2-\frac{r^2-1}{4t^2}
J_{2r+1}(4t)J_{2r-1}(4t)\,,
\ee
which is valid both for $h_0=0,\infty$.
Neglecting fast oscillations, $\langle \s_r^z(t)\s_0^z(t)\rangle_c$
increases as $r^2$ for $r<2t$ and then it drops almost immediately to $0$.
\item The autocorrelation function (not connected) of $\sigma_z$ at different
times is
\be
\langle \s_0^z(t_1)\s_0^z(t_2)\rangle
=J_0^2(2t_2-2t_1)-\frac{1}{4}[f(t_2+t_1)\pm g(t_2-t_1)]\,,
\ee
where $f(x)=J_2(2x)+J_0(2x)$, $g(x)=J_2(2x)-J_0(2x)$ and the sign $+$ ($-$)
refers to $h_0=0$ ($h_0=\infty$).
\end{itemize}

Let us comment on these results in view of the general
understanding we found so far. For $t<t^*$ all the connected
correlation functions are zero, in agreement with CFT.
All the oscillation terms of the asymptotic form $\cos (4t)$ are, as we
discussed in the previous section, a lattice effect.
Concerning the $\sigma_z$ correlator, the $r^2$ dependence
for $t>r/2$ is a consequence of the fact that $\sigma_z$ is not primary.
The same is true for the two-time correlations function of $\sigma_z$ that
also decays as a power law of $t-s$ for large times, instead of the
exponential prediction by CFT for primary field.

\section{Higher dimensions}
\label{HD}

Until now we just considered one-dimensional systems. Despite the
fact that in low dimensions the effect of fluctuations is more
pronounced making the physics highly non-trivial it is desirable
to have results in higher dimensions as well. The method presented
in Sec. \ref{secpath} to obtain the non-equilibrium dynamics of a
quantum model close to a critical point from the critical behavior
of a system confined in a slab geometry applies to generic
dimension $d$ through the study of the hamiltonian (\ref{LGH}).
Its analysis proceeds via field-theoretical RG that may provide
the all scaling quantities of the model in an expansion close to
the upper critical dimensions (u.c.d.), that is $D=d+1=4$. Above
the u.c.d. mean-field (or gaussian) results are exact, with
logarithmic correction at the u.c.d.. For dimensions lower than
the u.c.d., the scaling quantities are obtained as series in
$\e=4-(d+1)$. Thus for the time-evolution problem the simple
mean-field solution represents an exact scaling result (a part log
corrections) for the physically relevant three-dimensional case.
An alternative method to attack analytically the hamiltonian
(\ref{LGH}) is to consider an $N$ component field $\phi$ and
taking the limit $N\to\infty$, but this will not be employed here.

\subsection{Dirichlet boundary conditions: the two-point function}

The $D$-dimensional slab geometry with Dirichlet boundary
conditions has been the subject of several investigations. The
two-point function has been calculated at the first order in $\e$
expansion in Ref. \cite{kd-99}. The gaussian two-point function,
with partial Fourier transform in the parallel directions reads
\cite{kd-99} \be G(p,z_1,z_2)=\frac{1}{2b}\left(
e^{-b|z_1-z_2|}-e^{-b(z_1+z_2)}+\frac{e^{-b(z_1-z_2)}+e^{-b(z_2-z_1)}-e^{-b(z_1+z_2)}-e^{b(z_1+z_2)}}{e^{2bL}-1}
\right)\,, \label{Ggau} \ee with $b=\sqrt{p^2+m^2}$. We are
interested in the case where $L=2\t0$, $z_1=z_2=\tau$ that we will
analytically continue to $\tau\to \t0+i t$, and for computational
simplicity we will restrict to the massless case $m=0$. Thus in
real space and imaginary time, we have ($p=|{\mathbf p}|$) \bea
G(r,\tau)&=& \int \frac{d^3p}{(2\pi)^3} e^{-i {\mathbf{p\cdot
r}}}\frac{1}{2p}\left(
1-e^{-2p\tau}+\frac{2(1-\cosh(2p\tau))}{e^{4p\t0}-1}\right)=\nonumber\\
&=& \frac{1}{(2\pi)^2} \int_0^\infty p^2 dp \frac{1}{2p}\left(
1-e^{-2p\tau}+\frac{2(1-\cosh(2p\tau))}{e^{4p\t0}-1}\right)
\int_{-1}^1 d(\cos\theta)e^{ipr \cos\theta}=\nonumber\\
&=&\frac{1}{(2\pi)^2 r} \int_0^\infty dp \sin{pr}\left(
1-e^{-2p\tau}+\frac{2(1-\cosh(2p\tau))}{e^{4p\t0}-1}\right)\,.
\label{Gr3D}
\eea
This integral can be performed by making the sum over all the residues coming
from the denominator $e^{4p\t0}-1$. The calculation is rather involved, but
the final result is very simple:
\be
G(r,\tau)=
\frac{1}{\pi r\t0}\,\frac{\coth(\pi r/4\t0) \sin^2(\pi\tau/2\t0)}{
\cosh(\pi r /2\t0)-\cos(\pi\tau/\t0)}\,.
\ee
Continuing to real time $\tau=\t0+i t$ we obtain
\be
G(r,t)= \frac{1}{\pi r\t0}\,\frac{\coth(\pi r/4\t0) \cosh^2(\pi t/2\t0)}{
\cosh(\pi r /2\t0)+\cosh(\pi t/\t0)}\,,
\ee
that for $t,r\gg \t0$ simplifies to
\be
G(r,t)\simeq \frac{1}{r\t0}\,
\frac{ e^{\pi t/\t0}}{ e^{\pi t/\t0}+ e^{\pi r/2\t0}}\propto
\begin{cases}
e^{\pi (t-r/2)/2\t0}/r &{\rm for }\; t<r/2\,,\\
1/r &{\rm for }\; t>r/2\,.
\end{cases}
\ee
We recall that this result in $d=3$ is exact a part log corrections.
Thus the basic structure of the two-point function in 3D is the
same as in 1D, with a characteristic time $t^*=r/2$.
Using the result in Ref. \cite{kd-99} it is in principle possible to calculate
the correlation functions for $d<3$ in the $\e$ expansion framework. However,
this requires the analytical continuation of complicated functions, resulting
in a quite cumbersome algebra, as we shall see in the following for a simpler
observable.

Eq. (\ref{Ggau}) can be used in principle to determine the
gaussian behavior in any $d<3$. Unfortunately the integral one
gets is not analytically tractable for $d\neq1,3$. A possible
strategy would be to perform the analytic continuation before of
the integral and then evaluate it through a saddle-point
approximation. It is straightforward to show that the result
obtained in this way is equivalent to what we discuss in section
\ref{rtsm} where we remand for the analysis of $1<d<3$.

\subsection{Dirichlet boundary conditions: a non-trivial one-point function}

In the case of Dirichlet boundary conditions, the order parameter profile in
the slab geometry is trivially vanishing.
However not all the one-point expectation values are zero.
Let us consider as a typical example the operator
${\cal O}=\phi^2$ that has been calculated at the first order in $\e$
expansion in Ref. \cite{ked-95}. It has the scaling form
\be
\langle {\cal O}(z,L)\rangle \simeq L^{-d+1/\nu} H(z/L)\,,
\ee
where $\nu$ is the correlation length exponent.
In $D=4$, the function $H(x)$ is \cite{ked-95}
\be
H(x)=\frac{\pi^2}{\sin^2(\pi x)}-\frac{\pi^2}{3}\,,
\ee
that, using $L=2\t0$ and continuing to $z=\tau=\t0+it$, leads to the real
time evolution
\be
\langle {\cal O}(t)\rangle -\langle {\cal O}(t=\infty)\rangle
\simeq \t0^{-2}\frac{\pi^2}{\cosh^2(\pi t/2\t0)}\sim
\t0^{-2} e^{-\pi t/2\t0}\,,
\ee
i.e. $\langle {\cal O}(t)\rangle$ approaches its asymptotic value
exponentially, as it as been found for all primary operators in $d=1$.

The one-loop result for the $O(N)$ model in $D=4-\e$ dimensions
is \cite{ked-95}
\be
H(x)=\left[\frac{\pi^2}{\sin^2(\pi x)}-\frac{\pi^2}{3}\right]
\left(1+\e\frac{N+2}{N+8}\ln \frac{\pi}{\sin\pi x}\right)-
\e [\zeta'(2,x)+\zeta'(2,1-x)]+{\rm const}\,,
\ee
where const stands for terms do not depend on $x$ and
$\zeta'(\alpha,x)=\partial_\alpha \zeta(\alpha,x)$ and $\zeta$ is the
generalized Riemann function.
This example clearly shows that within the $\e$ expansion, the scaling
functions contain logarithmic contributions that originate from the expansion
of power law as e.g. $\sin^{\e} \pi x=1+\e\ln\sin \pi x$.
To have a function with good analytical structure to perform the real time
continuation, it is desirable to ``exponentiate'' such logarithms.
In Ref. \cite{ked-95} the exponentiation procedure leaded to
\begin{multline}
H(x)=\left(\frac{\pi}{\sin(\pi x)}\right)^{2-1/\nu}
[\zeta(D-2,x)+\zeta(D-2,1-x)-2\zeta(D-2)]\\-
\frac{\pi^2}{6}(2-1/\nu) \left(\frac{\pi}{\sin(\pi x)}\right)^{D-2-1/\nu}\,,
\end{multline}
with $\nu=1/2 +(N+2)/(N+8) \,\e/4$.
Performing the analytical continuation we have the sum of
two exponentials, and in any dimension $d<3$ the second one has a largest
``relaxation time'' that hence is dominating for large $t$.
Considering only the second term we have
\be
\langle {\cal O}(t\gg\t0)\rangle-\langle {\cal O}(t=\infty)\rangle
\propto e^{-(d-1-1/\nu)\pi t/2\t0}\,.
\ee
Note that the subleading term $ e^{-(2-1/\nu)\pi t/2\t0}$ is multiplied by a
$\log t$ term arising from the $\zeta$ function.

This example put forward the idea that (at least for Dirichlet boundary
conditions) the exponential relaxation of the one-point functions is not only
a property of
one-dimensional systems, but holds in any dimension (with eventually
$\log$ corrections) with relaxation times related to the scaling
dimensions of the operator.

\subsection{Fixed boundary conditions: the one-point function}

The case of the extraordinary transition, that corresponds to fixed boundary
conditions, has been considered in Ref. \cite{k-97}.
The magnetization profile can be written as \cite{k-97}
\be
\phi(z)=\frac{2K_4}{L} \frac{{\rm dn}(2K_4 z/L)}{{\rm sn}(2K_4 z/L)}\,,
\ee
where $K_4=K(k_4)$, where $K(k)$ is the elliptic integral, $k_4$ the elliptic
modulus that in terms of the parameter of the model is
$\phi^2 L^2=(2K_4)^2(2 k_4^2-1)$, and sn$(x)$ and dn$(x)$ and the Jacobi
functions.
Continuing to real time, and using the properties of the Jacobi functions
we obtain
\be
\phi(t)=\sqrt{1-k_4} {\rm cn} (K_4 t/\t0,1-k_4)\,,
\ee
that, contrarily to all the other cases we have considered, is
oscillating and not exponentially decreasing.

To our knowledge there are no result in the $\e$-expansion for the
magnetization profile. However, using ``local-functional methods''
\cite{bu-98} it has been obtained an approximate profile in $D=3$
that involves, as in mean-field theory, Jacobi elliptic functions.
It can be easily continued to real time via $z\to \t0+i t$ and again
one finds an oscillating behavior with time. The method exploited in
Ref. \cite{bu-98} can be used in any dimensions, obtaining an always
an oscillating behavior with a period that diverges as $D$
approaches 2, recovering Eq. (\ref{phistrip}).
All these calculations are essentially mean field and we do not know how
the inclusion of fluctuations changes them. It can possible that
for the extraordinary transition (i.e. fixed initial conditions), the
exponential decay founds at $D=2$ is more an exception rather than a rule,
because of the simple analytic structure of the trigonometric functions in the
complex plane. Another possibility is that fluctuations destroy these
oscillations. Only a complete analytical calculation (e.g. in large $N$)
can help in understanding this point.


\subsection{A real-time solvable model}
\label{rtsm}

As for the one-dimensional case, it is worth to check the results
coming from the analytical continuation of large imaginary-time
with exactly solvable models. The simplest (and probably one of
the few) model solvable in generic dimension is the generalization
of the hamiltonian (\ref{Hchaingen}) to a $d$-dimensional
hypercubic lattice. The solution of such model proceeds via
Fourier transform as in one dimension. The final result is simply
give by Eq. (\ref{2ptgaussian}) with the replacement $dk\to d^dk$,
i.e. \be \label{2ptgaussiand} \langle\p_r(t)\p_0(t)\rangle-
\langle\p_r(0)\p_0(0)\rangle =\int_{\rm BZ}
\frac{d^dk}{(2\pi)^d}e^{i\bf{k \cdot r}}
\frac{(\Omega_{0k}^2-\Omega_k^2)(1-\cos(2\Omega_kt))}{\Omega_k^2\Omega_{0k}}\,.
\ee First of all let us note that this expression in \em virtually
\em identical to Eq. (\ref{Gr3D}) if we take $\t0\propto
m_0^{-1}\to 0$. In fact, taking $\t0\to 0$ in Eq. (\ref{Gr3D}),
only the third term matters, since it is $O(\t0^{-1})$ relative to
the first two. Taking $z=r+it$ and $b=\Omega_p$ we get $G(p)\propto
\t0^{-1}\int \Omega_p^{-2} (1-\cos2\Omega_p t)$ that is Eq.
(\ref{2ptgaussiand}) for $m_0\to\infty$.

To understand the general features let us consider in details the conformal
evolution ($\Omega_k=v |{\bf k}|$) from a disordered state ($\Omega_{0k}= m_0$).
The derivative of the two-point function is ($k=|{\bf k}|$)
\be
\partial_t \langle\p_r(t)\p_0(t)\rangle=2m_0 {\rm Im}
\int \frac{d^dk}{(2\pi)^d} \frac{e^{i({\bf k \cdot r}-2k v
t)}}{k}\,. \ee Except for $d=1$, this can be done analytically
only in $d=3$, where we can write it as \be \sim \int k dk \int
d\theta \sin\theta e^{ik(r \cos\theta-2v t)}\sim \int dk
\big(\sin(kr)/r\big) e^{2ikvt}\sim (1/r) \delta(vt-r/2)\,. \ee
Integrating with respect to $t$ we get zero for $t<r/2v$ and $1/r$
for $t>r/2v$.

For general $d$ we have
\be
\partial_t \langle\p_r(t)\p_0(t)\rangle\propto 2m_0 {\rm Im} \int k^{d-2}dk\int
d\theta (\sin \theta)^{d-2}   e^{ik(r\cos\theta-2vt)}\,. \ee By a
saddle point argument, we can assume that the dominant behavior
comes from $\theta$ close to zero, so the $\theta$ integral gives
\be \int d\theta \theta^{d-2}e^{-ikr\theta^2/2}\sim
(kr)^{-(d-1)/2}\,, \ee leading to \be \int dk
\frac{k^{d-2}}{(kr)^{(d-1)/2}} e^{i(k(r-2vt))}\sim
r^{-(d-1)/2}(2vt-r)^{-(d-1)/2}\Theta(vt-r/2)\,. \ee Integrating
with respect to $t$, we get zero for $t<r/2v$, as expected by
causality, and \be r^{-(d-1)/2} (2vt-r)^{(3-d)/2}\,, \ee for
$t>r/2v$. It is interesting that this gaussian correlation
function blows up at large $t$ only for $d<3$, when we expect the
fluctuations to become important. It would be interesting to study
this in the $\phi^4$ theory for large $N$, by replacing $\phi^4$
by $3\langle\phi^2\rangle\phi^2$ in the usual way, where now
$\langle\phi^2\rangle$ depends on $t$ and is calculated
self-consistently.

\section{Physical interpretation and discussion}
\label{phint}

In this paper we studied in general the non-equilibrium unitary
dynamics that follows a sudden quantum quench. We showed that if
the hamiltonian $H$ governing the time evolution is at a critical
point, while $H_0$ (i.e. the one for $t<0$) is not, the
expectation value of a class of operators (primary ones in CFT)
relaxes to the ground-state value exponentially in time with
universal ratio of decaying constants. We also found that
connected two-point functions of operators at distance $r$ are
vanishing for $t<r/2v$, while for $t>r/2v$ reach exponentially
fast a value that depends \em exponentially \em on the separation,
in contrast with the power laws typical of equilibrium
configuration.

We also considered the real-time dynamics of simple exactly
solvable models and we found that several of the typical
characteristics of the critical points still hold. Roughly
speaking, critical points are not special as far as quenching
dynamics is concerned. In fact, also for gapped systems, connected
correlation functions vanish (or are strongly suppressed) for
$t<r/2v$ and for asymptotic large times resemble those at finite
temperature despite the fact that the whole system is in a pure
state. Several other examples in the recent literature (see e.g.
\cite{ir-00,cc-05,dmcf-06,c-06,gg,cdeo-06}) gives further evidence
that these two effects are actually true in general, at least in
the realm of exactly solvable models considered so far.

In the following we give a simple interpretation of  these two features
separately trying to understanding their physical origin.

\subsection{The horizon effect}

The qualitative, and many of the quantitative, features found for the time
evolution of correlation functions may be understood physically on the basis
of a picture we first introduced in Ref. \cite{cc-05} to describe the
time evolution of the entanglement entropy.
Later we generalized it to correlation functions in Ref. \cite{cc-06} and it
has been largely adopted thereafter \cite{dmcf-06,c-06,spinor,cdeo-06}.

We emphasize that such scheme is not an ab initio calculation but
rather a simplified picture which allows us to explain physically
our findings. The initial state $|\psi_0\rangle$ has an
(extensively) high energy relative to the ground state of the
hamiltonian $H$ which governs the subsequent time evolution, and
therefore acts as a source of quasiparticle excitations. Those
quasi-particles originating from closely separated points (roughly
within the correlation length $\xi_0=m_0^{-1}$ of the ground state
of $H_0$) are quantum entangled and particles emitted from far
different points are incoherent. If the quasiparticle dispersion
relation is $E=\Omega_k$, the classical velocity is ${\bf
v}_k=\nabla_k\Omega_k$. We assume that there is a maximum allowed
speed $v_m=\max_{\bf k} |{\bf v_k}|$. A quasiparticle of momentum
$k$ produced at ${\bf r}$ is therefore at ${\bf r}+{\bf v}_k t$ at
time $t$, ignoring scattering effects. This is the only physical
assumption of the argument. Scattering effects are not present in
the theories considered so far, but as evident from the argument
outlined below they can play a role for only $t>r/2v_m$ (allowing,
perhaps, for a renormalization of $v_m$ by the interactions).

These \em free \em quasi-particles have two distinct effects.
Firstly, incoherent quasi-particles arriving a given point $\bf r$
from well-separated sources cause relaxation of (most) local
observables at $\bf r$ towards their ground state expectation
values. (An exception is the local energy density which of course
is conserved.) Secondly, entangled quasi-particles arriving at the
same time $t$ at points with separation $|{\bf r}|\gg\xi_0$ induce
quantum correlations between local observables. In the case where
they travel at a unique speed $v$ (as in CFT), therefore, there is
a sharp ``horizon'' or light-cone effect: the connected
correlations do not change from their initial values until time
$t\sim |{\bf r}|/2v$. In the CFT case this horizon effect is
rounded off in a (calculable) manner over the region $t-|{\bf
r}|/2v\sim\tau_0$, since quasi-particles remain entangled over
this distance scale. After this they rapidly saturate to \em
time-independent \em values. For large separations (but still $\ll
2vt$), these decay \em exponentially \em $\sim\exp(-\pi x|{\bf r}|
/2\tau_0)$. Thus, while the generic one-point functions relax to
their ground-state values (we recall in CFT this relaxation is
exponential $\sim\exp(-\pi xvt/\tau_0)$), the correlation
functions do not, because, at quantum criticality, these would
have a power law dependence. Of course, this is to be expected
since the mean energy is much higher than that of the ground
state, and it does not relax.

This simple argument also explains why for the case of a semi-infinite chain
the relevant time scale is $r/v$ rather than $r/2v$, since one of the
two particles arriving in $r$ has been reflected from the end of the chain.
This has been also stressed in Ref. \cite{dmcf-06}, in the study of the time
dependence of entanglement entropy of finite chains with open boundary
conditions.

All our results are consistent with this picture as long as the
quasi-particles are assumed to all propagate at the same speed,
resulting from a ``conformal'' dispersion relation $\Omega_k=v
|k|$. However, it is very simple to generalize this picture to
different dispersion relations, taking into account that each
particle propagates at group velocity $v_k\equiv\Omega'_k$
appropriate to the wave number $k$. In this case the horizon
effect first occurs at time $t\sim |{\bf r}|/2v_m$, where $v_m$ is
the maximum group velocity.  If $v_m$ occurs at a non-zero wave
number, it gives rise to spatial oscillations in the correlation
function. Thus again connected correlation function are expected
to be strongly suppressed for $t<t^*$ and start developing only
after $t^*$. In the case of a general dispersion relation we do
not have a proof, beyond the stationary phase approximation, that
the connected correlation functions remain constant up to this
time, but one would expect it on the grounds of causality. (The
proof in the case of a relativistic dispersion relation uses
Lorentz invariance in an essential way.) However, because there
are also quasi-particles moving at speeds less than $v_m$, the
approach to the asymptotic behavior at late times is less abrupt.
In fact, for a lattice dispersion relation where $\Omega'_k$
vanishes at the zone boundary, the approach to the limit is slow,
as an inverse power of $t$. A similar result applies to the
$1$-point functions. This is consistent with the exact results
obtained here and elsewhere.

\subsection{The large time limit and the generalized Gibbs ensemble}

The existence and the understanding of the asymptotic state
resulting from the evolution from an arbitrary state is one of the
most-interesting problem in statistical mechanics. A robust theory
able to predict this state ab-initio still does not exist. A
currently popular idea is that for late times the system (or
rather macroscopically large subsystems) `look like' they are in a
thermal state, despite the fact that the actual state of the whole
system is pure. A common belief is that a region of dimension $r$
can be thermalized by the infinitely large rest of the system
which acts as a bath (see e.g. \cite{cc-05,eo-06,cdeo-06}). But
this intriguing idea is not sufficient to give the value of the
resulting effective temperature.

A major step toward the clarification of the properties of the
asymptotic state has been made by Rigol et al. \cite{gg}. In fact,
it was conjectured that if the asymptotic stationary state exists,
it is given by a generalized Gibbs ensemble obtained by maximizing
the entropy $S=-\Tr \rho \log \rho$, subject to all the
constraints imposed by the dynamics \cite{gg}. Consequently,
denoting  with $I_m$ a maximal set of commuting and linearly
independent integrals of motion, the density matrix is
\begin{equation}
\rho=Z^{-1} e^{-\sum_m \l_m I_m},\qquad Z= \Tr\, e^{-\sum_m \l_m I_m}.
\label{gge}
\end{equation}
We note that such a density matrix describes a pure state only if
the model under consideration is integrable, i.e. if the number of
integral of motions equals the number degrees of freedom. If there
are not enough integrals of motion, $\rho$ corresponds to a mixed
state and it is not clear to us to which extent it can describe
the pure state resulting from the time-evolution. The values of
the Lagrange multipliers $\l_m$ are fixed by the initial
conditions: \be \Tr\,I_m \rho= \langle I_m\rangle_{t=0}\,. \ee In
the following we will take this generalized Gibbs ensemble as a
postulate and we will show how it nicely and naturally explains
the ``effective temperature'' effect observed for large times.
However we stress that there is still no proof for this assumption
that, to our opinion, cannot be considered on the same fundamental
level as the thermal Gibbs ensemble.

Let us consider the chain of harmonic oscillators of the previous
section as a typical example. We will soon see that most of the
features are quite general. In this case the natural choice for an
infinite set of integral of motion is the number of particles with
momentum $k$, i.e. $n_k=A^\dag_k A_k$. Most other observables can
be written in terms of these, i.e. $H=\sum_k \Omega_k n_k$.
Consequently the expectation value is given by \be \langle {\cal
O} \rangle_{t=\infty}= \Tr {\cal O} \rho= \Tr {\cal O} Z^{-1}
e^{-\sum_k \l_k n_k}\,, \ee that can be seen as a thermal density
matrix with a $k$ dependent effective temperature given by \be
\b_{\rm eff}(k)\Omega_k = \l_k\,. \ee Thus an effective temperature
already appeared. Note that this state can still be pure because
such a temperature is $k$ dependent.

To fix $\l_k$ we need $\langle n_k\rangle_{t=0}$.
From Eq. (\ref{dk}) we get
\be
\langle n_k\rangle_{t=0}=
\langle A^\dag_k A_k\rangle_{t=0}=d_k^2=
\frac14\left(\frac{\Omega_k}{\Omega_{0k}}+\frac{\Omega_{0k}}{\Omega_k}\right)-\frac12\,.
\ee
The calculation then proceeds as for a thermal distribution
\be
\langle n_k\rangle_\rho= \Tr n_k \rho=-\frac{\partial}{\partial\l_k} \ln Z\,,
\qquad
{\rm with}
\quad
Z=  \Tr e^{-\sum_k \l_k n_k}=\prod_k \sum_{n_k=0}^\infty e^{-\l_k n_k}=
\prod_k\frac1{1-e^{-\l_k}}\,,
\ee
so that
\be
\langle n_k\rangle_\rho= \frac{\partial}{\partial\l_k}\sum_k \ln
(1-e^{-\l_k})=
\frac1{e^{\l_k}-1}\,.
\ee
From this $e^{\l_k}=1+d_k^{-2}$ and
\be
\beta_{\rm eff}(k)= \frac{\ln (1+d_k^{-2})}{\Omega_k}\,.
\ee

At finite temperature the correlation function in momentum space is
\be
\langle \p_k \p_{-k}\rangle_\b= \langle\frac2{\Omega_k} (A_k+A^\dag_k)
(A_{-k}+A^\dag_{-k})\rangle_\b=
\frac2{\Omega_k}\left(\langle A^\dag_k A_{-k} \rangle_\b
+\langle A_kA^\dag_{-k}\rangle_\b\right)=
\frac2{\Omega_k}\frac{1+e^{-\b \Omega_k}}{1-e^{-\b \Omega_k}}\,,
\ee
that substituting the previous result for $\b_{\rm eff}$ gives
\be
\rho(k)_{t=\infty}=\langle \p_k \p_{-k}\rangle_{t=\infty}=
\frac2{\Omega_k}(1+2d_k^2)=\frac{\Omega_k^2+\Omega_{0k}^2}{\Omega_k^2\Omega_{0k}}\,,
\ee
that is exactly the time-average of Eq. (\ref{MDF}) reproducing, after Fourier
transforming, the well defined correlation in real space for $t\to\infty$.

So this generalized Gibbs ensemble correctly reproduces the exact
diagonalization of the model and gives also few insights more.
In fact, in the limit $m_0\to\infty$ the effective temperature is
independent from $k$ and $m$ obtaining  $\b_{\rm eff}=4/m_0$, explaining a
posteriori the simplicity of the results in this case.
Note instead that for arbitrary $m_0$ and $m=0$, i.e. conformal evolution,
$\b_{\rm eff}(k)$ is a function of $k$. In this case, the large distance
properties of correlation functions are described by the mode with $k=0$, for
which, independently of $m_0$ we get $\b_{\rm eff}(k=0,m=0)=4/m_0$,
consistently with the previous findings and general expectations.
Furthermore we can conclude that the large $r$ asymptotic behavior is always
governed by the effective temperature
$\b_{\rm eff}(k=0)=2(\log(|m_0-m|/(m_0+m))/m$.
Another interesting feature is that $\b_{\rm eff}(k=0)$ gives the asymptotic
behavior of the correlation functions of all those observables that are
effectively coupled with the zero-mode. In the opposite case the relevant
temperature is the largest $\b_{\rm eff}(k)$ with the $k$ mode coupled with
the observable.
Another interesting feature is that on a very rough basis one can be tempted to
assume that that $\b_{\rm eff}(k)$ is directly related to the excess of energy
of the mode $k$ generated by the quench, but this is not the case.
In fact $\langle H\rangle_{t=0}=\sum_k \Omega_k d_k^2$ that is different from
$\b_{\rm eff}^{-1}$, being the same only in the limit $d_k^{-2}\to0$,
i.e. $m_0\to\infty$.

Clearly the same reasoning of before applies every time we consider a model
that can be diagonalized in momentum space with a proper choice of the
quasi-particles, i.e. every time that $H=\sum_k \Omega_k A_k^\dag A_k$ for some
$A_k$. This means that the excitations are non-interacting.
As far as we are aware all the applications of the generalized Gibbs ensemble
to the date only concern this kind of models \cite{gg,c-06}, but there are
few numerical hints suggesting that can
be true more generically \cite{mwnm-06}.

Now we outline how we imagine the generalized Gibbs ensemble given
by Eq. (\ref{gge}) could be used to justify the effective
temperature scenario for large time for any integrable system,
i.e. with a complete set of integrals of motion. The hamiltonian
can be written as $H=\sum_m a_m I_m$, with some $a_m$ eventually
zero. Thus one can think to an $m$-dependent effective temperature
$\beta_{\rm eff}(m)=\l_m/a_m$, but this temperature does not give
directly the behavior of the correlation function for large
distance because in general the integral of motions are not
diagonal in $k$-space.
Thus we can only conjecture that the correlation functions of a
given operator $O(r)$ are governed by the largest
 $\beta_{\rm eff}(m)$ with $m$ among the integral of
motions to which $O(r)$ effectively couples. We stress that this is only a
crude argument and we are still not able to put it on a firmer basis.

\subsection{Discussion and open questions}

We presented a quite complete picture on the time evolution of a
quantum system after a sudden quench of one hamiltonian parameter.
Despite the fact that a lot of work has been done, still more is
left for future investigation.

The first problem that must be addressed is the real-time dynamics
of effectively interacting systems.
In this direction Bethe Ansatz solvable models like
the Lieb-Liniger gas and Heisenberg spin chains are among
the best candidates for
an analytical approach. This would clarify how a non-trivial two-body
scattering matrix can modify the time-evolution (only inside the light-cone)
of the correlation functions and whether the crude argument we outlined for an
``effective temperature'' for the long-time state is valid.
To clarify this point also
the study of (boundary) integrable massive field theories can be of some help.

Another approach, currently under investigation\cite{sotcardy}, is
the direct analysis of the perturbative expansion for the
correlation functions in a $\lambda\phi^4$ field theory. This has
some simplifying features in the large $m_0$ limit which may allow
it to be resummed to all orders. It is very important to get
non-trivial results for such a non-integrable interacting theory.

Numerical computations for non-integrable systems should also be performed to
understand if and eventually to which extent the generalized Gibbs ensemble
picture is valid beyond integrability. Some numerics concerning this point are
already available \cite{ksdz-05,kla,mwnm-06} but still a clear scenario is not
emerged.
Among non-integrable models special care must be given to disordered systems
because of the non existence of a speed of sound as a consequence
of Anderson localization \cite{bo-07}.
Some insights can be obtained from those models whose equilibrium behavior
can be analytically obtained by means of the strong-disorder
renormalization-group \cite{sdrg}.
Even in this case time-dependent DMRG can help a general understanding.
The time-evolution of the entanglement entropy \cite{dmcf-06} already revealed
that in these systems there is no sign of the light-cone (as easily predictable
because of the absence of speed of sound) and the effective motion of
``quasi-particles'' is more diffusive rather than ballistic. Furthermore the
results at the largest available times does not look thermal at all, but this
can be also due to the time window accessible with numerics.

Another very interesting question is what happens at finite
temperature and under what conditions a system can equilibrate. It
should be relatively simple to generalize the results for all the
``quasi-free'' models already considered at $T=0$. Work in this
direction is in progress, but it will be almost impossible to
consider the same problem for more complicated models.

Finally it is also important to understand the role played by the
initial state. We assumed always a translational invariant one
with short range correlations. Thus one natural modification
consists in taking a state that is only locally different from the
actual ground-state. This problem is known in the literature as a
``local quench'' and it has already been considered to some extent
(see e.g. \cite{aop-03,ksz-05,ep-07}), but still a general picture
as the one outlined here for global quenches does not exist (this
is also interesting for eventual connections with quantum impurity
problems, see \cite{s-98} as a review). Another natural
modification of the initial state is one with long-range
correlations, such as a critical one and let it evolve with
another critical hamiltonian. This has been done for the Luttinger
liquid \cite{c-06} and the results show the typical functional
dependence of a light-cone scenario (i.e. everything depends only
on $x\pm 2vt$) but the long-time correlations decays as
power-laws, with exponents that are different from equilibrium
ones and can be predicted by the generalized Gibbs ensemble.

\section*{Acknowledgments}
This work was supported in part by EPSRC grants GR/R83712/01 and
EP/D050952/1.
This work has been finished when PC was a guest of the Institute for
Theoretical Physics of the Universiteit van Amsterdam.
This stay was supported by the ESF Exchange Grant 1311 of the INSTANS activity.

\end{document}